# An Algorithm Framework for the Exact Solution and Improved Approximation of the Maximum Weighted Independent Set Problem

Kai Sun*

**Abstract**

The Maximum Weighted Independent Set (MWIS) problem, which considers a graph with weights assigned to nodes and seeks to discover the "heaviest" independent set, that is, a set of nodes with maximum total weight so that no two nodes in the set are connected by an edge. The MWIS problem arises in many application domains, including the resource-constrained scheduling, error-correcting coding, complex system analysis and optimization, and communication networks. Since solving the MWIS problem is the core function for finding the optimum solution of our graph-based formulation of the resource-constrained Process Planning and Scheduling (PPS) problem, it is essential to have "good-performance" algorithms to solve the MWIS problem. In this paper, we propose a Novel Hybrid Heuristic Algorithm (NHHA) framework in a divide-and-conquer structure that yields optimum feasible solutions to the MWIS problem. The NHHA framework is optimized to minimize the recurrence. Using the NHHA framework, we also solve the All Maximal Independent Sets Listing (AMISL) problem, which can be seen as the subproblem of the MWIS problem. Moreover, building composed MWIS algorithms that utilizing fast approximation algorithms with the NHHA framework is an effective way to improve the accuracy of approximation MWIS algorithms (e.g., GWMIN and GWMIN2 (Sakai et al., 2003)). Eight algorithms for the MWIS problem, the exact MWIS algorithm, the AMISL algorithm, two approximation algorithms from the literature, and four composed algorithms, are applied and tested for solving the graph-based formulation of the resource-constrained PPS problem to evaluate the scalability, accuracy, and robustness.

## 1 Introduction

As one of the most challenging problems in graph theory, the problem of finding the Maximum Weighted Independent Set (MWIS) considers a graph with weights assigned to nodes and seeks to identify the "heaviest" independent set, that is, a set of nodes with maximum total weight so that no two nodes in the set are connected by an edge. We name such a graph as a conflicting weighted graph. The statement of the MWIS problem looks relatively simple; however, solving the MWIS problem on general graphs is computationally difficult. It has been shown to be an NP-hard problem (Köhler & Mouatadid, 2016), so it is unlikely to be solved in polynomial time.

Inspired by a simple philosophy understanding, "nothing in the world stands by itself. Every object is connected in an endless chain and is thus connected with all the other links. And this chain unites all objects and processes in a single whole and thus has a universal character," if we can abstract a real-world engineering/optimization problem as a graph with nodes representing subproblems (or possible solutions of subproblems) and edges representing their relationships, then the global optimum solution can be found by solving the MWIS problem on such a graph. This type of formulation arises in many applications, including resource allocation, scheduling and staffing problems, error-correcting coding, complex system analysis and optimization, logistics and transportation, and communication networks. Modeling scheduling problems as such graphs are particularly relevant in the presence of incompatible entities to be scheduled. Our particular interest is focusing on the general type of Process Planning and Scheduling (PPS) problem (Sun et al., preprint). Therefore, in order to utilize the concept of the MWIS in our PPS application, low-complexity algorithms for solving the MWIS problem that yields "good-quality" feasible solutions (guaranteed independent nodes in the set) are desired.

*Department of Mechanical & Aerospace Engineering, College of Engineering & Computer Science, Syracuse University, Syracuse, NY 13244, USA. (kasun@syr.edu)

In this paper, we propose a Novel Hybrid Heuristic Algorithm (NHHA) framework in a divide-and-conquer structure that yields optimum feasible solutions to the MWIS problem. The NHHA framework is optimized to minimize the recurrence with a computational complexity that close to the pioneering work (Moon & Moser, 1965) whose complexity is $O(3^{\frac{n}{3}})$ on an n-vertex general graph. Using the NHHA framework, we also solve the All Maximal Independent Sets listing (AMISL) problem, which can be seen as the subproblem of the MWIS problem. Moreover, based on the NHHA framework, we developed two merging methods to compose the exact MWIS algorithm with fast approximation MWIS algorithms for faster computational speed or improved accuracy. All eight algorithms for the MWIS problem, the exact MWIS algorithm, the AMISL algorithm, two approximation algorithms (GWMIN and GWMIN2 (Sakai et al., 2003)), and four composed algorithms, are applied and tested for solving the graph-based formulation of the PPS problem (Sun et al., preprint). Note that this is a preprint of the paper that is scheduled to submit to Applied Discrete Mathematics.

The remainder of the paper is organized as follows. Section 2 provides a comprehensive literature review on the MWIS problem and its applications in scheduling problems. Section 3 provides the necessary background and definitions of graph theory. Section 4 and Section 5 explain the NHHA framework in detail, and Appendix I illustrates the exact MWIS algorithm with a simple example. Section 6 discusses how to merge the proposed MWIS algorithm with approximation MWIS algorithms to reduce the complexity. Then, Section 7 presents the numerical results to assess the performance of the algorithms. Finally, a few concluding remarks are given in Section 8.

## 2 Literature Review

There has been extensive work in the literature proposing a variety of algorithms for solving the MWIS problem exactly or approximately. One brute-force algorithm for exactly solving the MWIS problem amounts to checking all **M**aximal **I**ndependent **S**ets (MIS) and picking one with the maximum total weight. It follows that the MWIS problem is converted to the **A**ll **M**aximal **I**ndependent **S**ets (AMIS) listing (AMISL) problem (or maximal cliques listing problem in the complement graph). A pioneering work (Moon & Moser, 1965) has shown that any n-vertex graph has at most $3^{\frac{n}{3}}$ maximum cliques. Many algorithms are now known for the clique (or independent set) listing problem (Bron & Kerbosch, 1973; Loukakis & Tsouros, 1981; Johnson et al., 1988; Makino & Uno, 2004; Eppstein, 2005; Tomita et al., 2006; Cazals & Karande, 2008). Among those algorithms, a simple recursive backtracking algorithm (Bron & Kerbosch, 1973), Bron-Kerbosch algorithm named after its inventors, has been reported as the most successful clique listing algorithm in practice (Eppstein et al., 2010).

Other than the costly non-polynomial algorithm for the optimum solution on general graphs, people naturally go to three types of solutions: (i) solutions for special cases, it is known to be solvable in polynomial time in many cases including perfect graphs (Grotschel et al., 1993), interval graphs (Grotschel et al., 1993), disk graphs (Matsui, 1998), claw-free graphs (Minty, 1980), fork-free graphs (Alekseev, 2004), trees (Chen et al., 1988), sparse random graphs (Karp & Sipser, 1981; Czygrinow & Hanckowiak, 2006), circle graphs (Valiente, 2003), and growth-bounded graphs (Gfeller & Vicari, 2007). The MWIS problem has been found to be solvable in strongly polynomial time only on perfect graphs and their complements, on t-perfect graphs, and on claw-free graphs (Schrijver, 2003). (ii) approximation algorithms, there has been extensive work on approximating the MWIS (Halldorsson, 2004). The approximation can be achieved by using a greedy strategy (Furer & Kasiviswanathan, 2007). Sakai et al. (Sakai et al., 2003) investigated the performance guarantee of greedy algorithms to solve the MWIS problem. And (iii) there has been extensive work in the literature proposing a variety of heuristics (Kako et al., 2005). These specialized or heuristics algorithms have been developed for computing the exact MWIS (Fomin et al., 2006; Babel, 1994; Ostergard, 2002; Tassiulas & Ephremides, 1992) for limited types of graphs or graphs in general with certain trade-offs.

The Graph Coloring Problem (GCP), MWIS, and AMISL problems, as well as their extensions, have been proposed to cope with different scheduling environments. The GCP consists of assigning a single color (integer) to each vertex of an undirected graph, such that no two adjacent vertices share the same color, intending to minimize the number of colors (Tucker, 2012). The MWIS problem is a special case of the GCP, when each node is associated with a weight factor with an optimization objective of finding the set maximizing the total weight for each coloring terms of finding the optimum set of nodes for each color. With interests focusing on scheduling problems, we summarize the four typical scheduling problem formulations with the GCP and its variations. Although these formulations are not fitting in the resource-constrained PPS problem considered in our work (Sun et al., preprint), but they are inspiring for us to develop our graph-based formulation.

(1) The Class/Exam Scheduling Problem

The class scheduling problem, also named as the timetabling problem, can be stated as follows: schedule a set of classes in a number of time slots such that no professor or student is required at the same time. Constraints can be mapped onto GCP as follows. Let each class be represented by a node. Attach two nodes by an edge if and only if there is a reason that the classes they represent may not be offered at the same time. Initially, there are two such reasons for nodes to be linked: either they are taught by the same instructor, or they are required by the same set of students. Upon adding in the links, color the graph. Each color represents a time slot available on a given timetable, so every node with the same color is offered at the same time. Similar applications of this problem can be the scheduling of classes and exams in a university, the scheduling of flights for an airline, and the scheduling of computing tasks to be run on a multiprocessor machine (Dandashi & Al-Mouhamed, 2010; Miner et al., 1995).

(2) The Interval Graph Scheduling

An interval graph is the intersection graph of a set of intervals of a real line, that is, a graph whose nodes correspond to intervals such that two nodes connected by an edge are associated with intersecting intervals, as shown in Figure 1 (Gardi, 2009). The intervals are representing the tasks, and the edges are indicating the incompatible tasks. The graph is then colored to find the mutual exclusion tasks that can be processed by the same resources. The interval graph scheduling and many variants of this problem have been extensively studied due to its numerous applications (Krarup & De Werra, 1982; Blazewicz et al., 2001; Zais & Laguna, 2016).

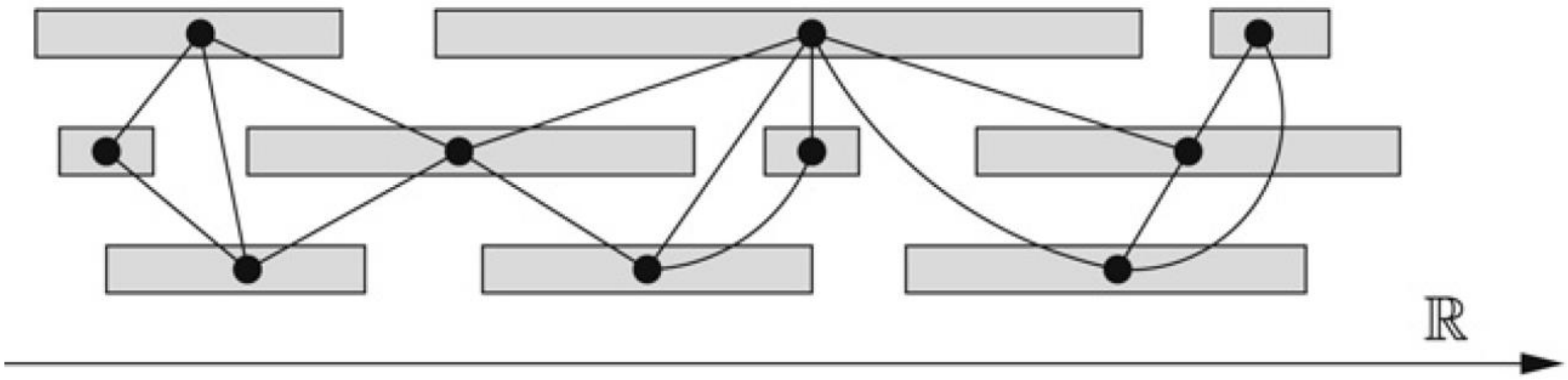

Figure 1. A Sample Interval Graph (Gardi, 2009)

(3) The Scheduling of Wireless Network

In a wireless network, two wireless nodes that transmit at the same resource (frequency), interfere with each other if they are located close-by. The scheduling problem is to decide which nodes should transmit at the given resource so that there is no interference, and nodes with longer queue length are given priority. If each node is given a weight equal to the queue length, it is optimum to schedule the set of nodes with the highest total weight. If a conflicting weighted graph is made, with an edge between each pair of interfering nodes, the scheduling problem is exactly the MWIS problem. This type of scheduling problem is mostly found in wireless communication applications (Tassiulas & Ephremides, 1992; Joo et al., 2013; Du & Zhang, 2016), but it is also applied in other types of applications (Duarte et al., 2015; Todosijevic & Mladenovic, 2016; Hansen et al., 2017; Gainanov et al., 2018).

(4) Graph Multi-coloring

The graph multi-coloring problem is an extension of the GCP. In this case, a node coloring corresponds to a sequence of colors (from the smallest to the largest). A node stands for a task, and an edge indicates that two tasks represented by the two end nodes of the edge are incompatible. Each color is a time slot, and each node must be assigned with a number of colors as defined by the processing time of the task. The objective is to minimize the number of used colors. Thevenin et al. applied this formulation in a flow production environment (Thevenin et al., 2018). The graph multi-coloring problem formulation is the closest formulation comparable to our PPS problem. Still, it can only be applied in restricted conditions, such as each job requires the resources continuously and no subtasks of each job.

# 3 Definitions and Notations

Let $G = (V, E)$ be a simple undirected graph with vertex set $V = \{1, \dots, v\}$, and a set of edges $E$. We denote by $|A|$ the cardinality of set $A$, so that the **edge number** of $G$ is $|E|$ and the **node number** of $G$ is $|V|$. Let $x \in V$, the **degree** (valence) of $x$ is the number of edges with $x$ as an endpoint. We denote the degree of $x$ by $d_G(x)$. Let $Neig_G(x)$ denote the set of neighbors of vertex $i$ and $Neig_G^+(x)$ denote $\{x\} \cup Neig_G(x)$. $d_G(x) = |Neig(x)|$ is the degree of vertex $x$.

In the graph $G$, let $S \subset V$ be any subset of vertices of $G$. Then, the induced subgraph $Ind_G(S)$ is the graph whose vertex set is $S$ and whose edge set consists of all of the edges in $E$ that have both endpoints in $S$ (Diestel, 2006). For a vertex $k \in V$, let the complementary induced subgraph $C_Ind_G(k)$ refers to the subgraph induced by all the node in $V$ except node $k$, and the complementary neighbor induced subgraph $C_Neig_Ind_G\ (k)$ refers to the subgraph induced by the non-neighbors of $k$, and $k$ is not in $C_Neig_Ind_G\ (k)$.

In the graph $G$, assume there is a sequence of vertices and edges $x_0, e_1, x_1, e_2, \ldots, e_n, x_n$, where, for all $i = 1, \ldots, n, x_{i-1}$ and $x_i$ are the endpoints of $e_i$ is called a **walk** ($x_0, x_n$-walk) in G from $x_0$ to $x_n$. A walk in which all edges are distinct is called a **trail** ($x_0, x_n$-trail) and a walk in which all vertices are edges are distinct is called a **path** ($x_0, x_n$-path). The length of this walk, trail, or path is $n$. The length of the shortest walk, trail, or path joining the vertex $x$ to the vertex $y$ is called the distance from $x$ to $y$.

A connected, acyclic (no circuits) graph is called a tree. The components of an arbitrary acyclic graph are trees, and an acyclic graph is called a forest.

In the graph $G$, a subset $I \subseteq V$ is called an **independent set** (stable set, vertex packing) if the edge set of the subgraph induced by $I$ is empty. An independent set is **maximal** (maximal independent set) if it is not a subset of any larger-size independent set, and **maximum** (maximum independent set) if there are no larger-size independent sets in the graph. The independence number $\alpha(G)$ (also called the stability number) is the cardinality of a maximum independent set in $G$. For each node $i \in V$, there is a positive weight $w_i > 0$. A subset of $V$ can be represented by binary variable $x_i$, $(1 \le i \le |V|)$, where $x_i$ is 1 if $i$ is in the subset and 0 otherwise. A subset is called an independent set if no two nodes in the subset are connected by an edge. We are interested in finding the MWIS (Papadimitriou and Steiglitz, 1982), which can be expressed as an integer program:

$$\max \quad \sum_i w_i x_i$$

$$s.t. \quad x_k + x_i \le 1, \qquad (k,i) \in E$$

$$x_i \in \{0,1\}, \ \ i \in V$$

# 4 The MWIS Algorithm Framework

The proposed NHHA framework has two phases following a divide and conquer structure: it starts by (a) removing nodes to get the induced subgraphs that are simple enough for finding the MWIS; and then by (b) iteratively adding nodes back one at a time, compare and merge to get the output. The first phase recursively partitions the graph into complementary induced subgraphs by removing nodes (and the adjunct edges) one at a time based on node removal heuristics. When induced subgraphs satisfy the desired patterns, these induced subgraphs become simple enough to be solved for MWIS with one comparison. A **Preliminary Set** (AMISL Preliminary Sets for the AMISL case) is found based on this complementary induced subgraph. The second phase of the algorithm adds back the nodes (and the adjunct edges) removed in the reversed sequence. At each adding, a **Compare Set** (AMISL sets for the AMISL case) is found to compare with the **Preliminary Set** (AMISL Preliminary Sets for the AMISL case). For the MWIS problem, the MWIS output set is the set with larger total weights among the Preliminary Set and the Compare Set of the current graph in the node adding process. For the AMISL problem, the AMISL output sets are the union of AMISL Compare Sets and AMISL Preliminary Sets for the graph with the adding node. The algorithm stops when all nodes (and the adjunct edges) are added back to the graph. With this brief understanding of the proposed approach, we are going into the details in the following sections.

**Phase I: Dividing**

Three types of unit graph structures (shown in Figure 2) are defined as **C**onnected **U**nit **S**ubstructures (CUS). The three types of CUS are: (a) an isolated node; (b) a pair of two connected nodes; and (c) a tree with a maximum diameter of 2 edges. Given an undirected weighted graph $\Gamma$ consists of $n$ different CUSs, $CUS_1, CUS_2, \ldots, CUS_i, \ldots, CUS_n, i \in \{1,2,\ldots,n\}$, and no edge between these CUSs. Define $MWIS(\Gamma)$ as a set of nodes, and this set has the maximum total weight in $\Gamma$. We denote the $MWIS(\Gamma)$ as the MWIS of graph $\Gamma$, $MWIS(\mathrm{CUS}_1)$, $MWIS(\mathrm{CUS}_2), \ldots, MWIS(\mathrm{CUS}_i), \ldots, MWIS(\mathrm{CUS}_n)$ as the MWISs of the CUSs, respectively. The $AMIS(\Gamma)$ is a set of all maximal independent sets in $\Gamma$. We denote the $AMIS(\Gamma)$ as the AMIS of graph $\Gamma$, $AMIS(CUS_1)$, $AMIS(CUS_2), \ldots, AMIS(CUS_i), \ldots, AMIS(CUS_n)$ as the AMIS of the CUSs, respectively. We denote the maximal independent set as $MIS_{CUS_i}^{k_i}$, which is an element in $AMIS(CUS_i) = \{MIS_{CUS_i}^{1}, MIS_{CUS_i}^{2}, \ldots, MIS_{CUS_i}^{k_i}, \ldots, MIS_{CUS_i}^{m_i}\}$, where $k_i \in \{1,2,\ldots,m_i\}$.

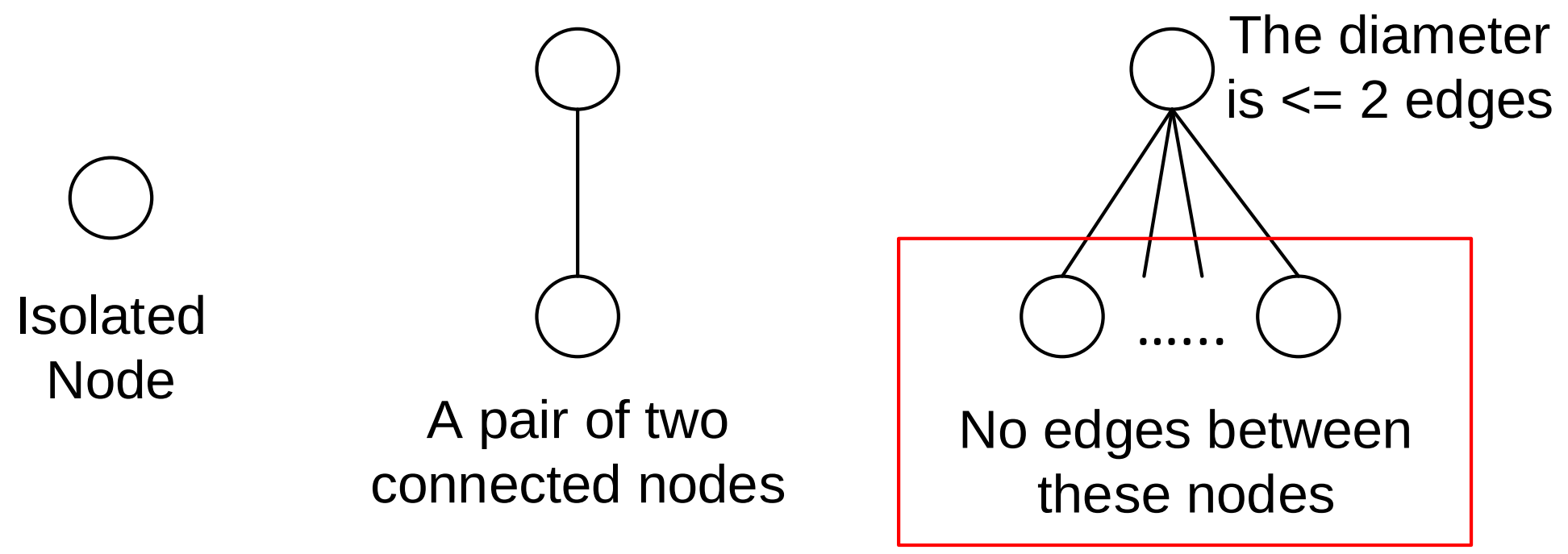

Figure 2. Three Types of Connected Unit Substructures (CUSs)

Theorem 1 and Corollary 1 show that we are able to find the MWIS and AMIS of an induced subgraph after partitioning it into a specific structure. In order to partition the graph to get an induced subgraph as $\Gamma$ described in Theorem 1, we need to proceed in two steps: (a) break all the cycles in the graph, and (b) break the paths which are longer than 2 edges. In both steps, we need to remove the nodes (and the adjunct edges) which satisfying specific rules. We denote such a qualified node as a **removed node**.

**Step 1**: Break all cycles

First, we need to find a cycle basis of the given graph $G = (V, E)$. For each node $i \in V$ in $G$, count the number of basic (fundamental) cycles it belongs to, we denote the count for node $i$ as $C_G(i)$. Then, we remove a node $n \in V$ to get the complementary induced subgraph $C_Ind_G(n)$, where $C_G(n)$ is the maximum among all the $C_G(i)$. This process iterates until no cycle left in the induced subgraph. This induced subgraph left is either a tree or a forest, since all the cycles are broken by removing the node (and adjunct edges) belongs to the most cycles.

**Theorem 1**: For Base Cases in Recurrence
Given a graph Γ that consists of $n$ different CUSs, $CUS_1$, $CUS_2$, …, $CUS_i$, …, $CUS_n$, $i \in \{1,2,\dots,n\}$, and no edge between these CUSs:

(i) For the MWIS problem, the $MWIS(\mathrm{CUS_i})$ can be found by one comparison in a $CUS_i$. The CUS with an isolated node can be considered as compared with an empty node set. For the Γ that consists of multiple CUSs, the $MWIS(\Gamma)$ is the union of the MWIS of each CUS in Γ, or formally,

$$MWIS(\Gamma) = MWIS(CUS_1) \cup MWIS(CUS_2) \cup \dots \cup MWIS(CUS_n)$$

(ii) For the AMISL problem, the $AMIS(CUS_i)$ can be found by dividing the graph into two independent node sets in a $CUS_i$. The CUS with an isolated node can be considered as dividing the graph into two node sets (one of the two sets can be an empty set, $\phi$). For the Γ that consists of multiple CUSs, each $CUS_i$ , $i \in \{1,2,\dots,n\}$ , in Γ has its $AMIS(CUS_i)$. The $AMIS(\Gamma)$ of graph Γ is all the combinations of the MISs of all the CUSs, note that only picking one of the MISs from each CUS in one combination. For the $AMIS(\Gamma)$ of graph Γ, $k_i \in \{1, 2, \dots, m_i\}$, formally,

$$AMIS(\Gamma) = \{ \{MIS^1_{CUS_1}, MIS^1_{CUS_2}, \dots, MIS^1_{CUS_n}\}, \dots, \{MIS^{k_1}_{CUS_1}, \dots, MIS^{k_i}_{CUS_i}, \dots, MIS^{k_n}_{CUS_n}\}, \dots, \{MIS^{m_1}_{CUS_1}, \dots, MIS^{m_i}_{CUS_i}, \dots, MIS^{m_n}_{CUS_n}\} \}$$

**Proof of Theorem 1:**

(i) For the MWIS problem, the MWIS can be found by one comparison in a CUS, because there are only two maximal independent sets in all the three types of CUSs. Since Γ consists of multiple CUSs, and there are no edges between these CUSs, the MWIS of each CUS does not have a conflict with the MWIS of another CUS in Γ. Because the MWIS of each CUS in Γ is the independent set with the possible maximum total weight, and MWISs of CUSs has no conflict with each other. We can get the union of MWISs of CUSs in Γ as the MWIS of Γ. ■

(ii) For the AMISL problem, the AMIS can be found by dividing a CUS into two independent node set. Since Γ consists of multiple CUSs, and there are no edges between these CUSs, the AMIS of each CUS does not have confliction with any node of another CUS in Γ. Because the AMISs of different CUSs in Γ has no confliction, get the union of the sets by choosing one set from the AMIS of each CUS and find all combinations without repeating of such unions. The union of each combination is one maximal independent set in the AMIS of Γ. ■

**Corollary 1:** The below statements in Theorem 1,
"For the $\Gamma$ that consists of multiple CUSs,

(i) For the MWIS problem, the $MWIS(\Gamma)$ is the union of the MWIS of each CUS in $\Gamma$, or formally,

$$MWIS(\Gamma) = MWIS(CUS_1) \cup MWIS(CUS_2) \cup \ldots \cup MWIS(CUS_n)$$

(ii) For the AMISL problem, the $AMIS(\Gamma)$ of graph $\Gamma$ is all the combinations of the MISs of all the CUSs, note that only picking one of the MISs from each CUS in one combination. For the $AMIS(\Gamma)$ of graph $\Gamma$, $k_i \in \{1, 2, \ldots, m_i\}$, formally,

$$AMIS(\Gamma) = \left\{ \begin{array}{c} \left\{MIS^{1}_{CUS_1}, MIS^{1}_{CUS_2}, \ldots, MIS^{1}_{CUS_n}\right\}, \\ \ldots, \\ \left\{MIS^{k_1}_{CUS_1}, \ldots, MIS^{k_i}_{CUS_i}, \ldots, MIS^{k_n}_{CUS_n}\right\}, \\ \ldots, \\ \left\{MIS^{m_1}_{CUS_1}, \ldots, MIS^{m_i}_{CUS_i}, \ldots, MIS^{m_n}_{CUS_n}\right\} \end{array} \right\}$$

",
also holds when the $CUS$ is a general graph.

**Proof of Corollary 1:** In Corollary 1, the CUS in $\Gamma$ in Theorem 1 is now a general graph. In other words, the connected components in $\Gamma$ is a general graph. Similar to the proof of Theorem 1, because the MWISs or AMISs of these connected components in $\Gamma$ has no conflict with nodes in a different connected component of $\Gamma$, so that Corollary 1 holds which means that Theorem 1 also holds when CUS is a general graph. ■

A basis for cycles of an undirected graph (**Cycle Basis**) is a minimal collection (a set of fundamental cycles) of cycles such that any cycle in the graph can be written as a sum of cycles in the Cycle Basis set (Diestel, 2012). Here summation of cycles is defined as "exclusive or" of the edges. The algorithm for finding a cycle basis is adapted from algorithm CACM 491, originally developed by K. Paton. For details on the algorithm and the production of the basic cycles, Paton's original paper (Paton, 1969) should be consulted. Paton also discusses two other algorithms for basic cycle generation and contains performance statistics in the paper referred to. The adopted basic (fundamental) cycles algorithm can be depicted as in Algorithm 1 (Paton, 1969).

**Step 2**: Break the paths which are longer than 2 edges to reduce the diameter of the components of the induced acyclic subgraph from step 1

If any of the connected components of the induced subgraph from step 1 has a diameter that is no less than 3 edges, remove the node in the middle of the longest path in that connected component of the graph. We name this node as the **Middle Node** of the path. For an odd path, the Middle Node is the midpoint of the path; for an even path, the Middle Node is one of the two nodes in the middle of the path. Algorithms 2 are adopted for checking the diameter, and Algorithms 3 is implemented for finding the Middle Node, respectively.

The diameter is the maximum eccentricity. The eccentricity of a node $v$ is the maximum distance from $v$ to all other nodes in $G$. If $G$ is disconnected, the eccentricity of a node $v$ is infinite. A diameter algorithm adapted based on the work by F. W. Takes, and his colleagues (Takes & Kosters, 2011; Takes & Kosters, 2013; Borassi et al., 2015) is applied here for computing the diameters in step 2. For each connected component of $G$, we utilize a function "**single_source_shortest_path_length**" from the python module "networkx" to compute the shortest path lengths from each node to all reachable nodes. The maximum value of the lengths found is the diameter of the connected component of $G$. We mark this algorithm as Algorithm 2, the diameter algorithm.

**Algorithm 1:** The Basic Cycles Algorithm
**Input:**
A graph is finite, connected, undirected, and without loops or multiple edges.
**Step 1:**
Let vertex 1 be the root of the spanning tree. Start forming the spanning tree by placing all edges of the form $\{1, W\}$ into the tree. At the same time, place all vertices W into a push-down list called STACK.
**Step 2:**
Let Z be the last vertex added to STACK (i.e. the top of the stack). If STACK is empty, then stop. If STACK is not empty, then remove Z from STACK and go to step 3.
**Step 3:**
Consider all edges $\{Z, W\}$ which have not been examined. If all edges have been examined, go to step 2. Otherwise, for each edge $\{Z, W\}$ do the following:

a. If W is in the tree generate the basic cycle formed by adding $\{Z, W\}$ to the tree and repeat step 3.
b. If W is not in the tree, add $\{Z, W\}$ to the tree, W to STACK, and repeat step 3.

Algorithm 1: The Basic Cycles Algorithm (Paton, 1969)

The diameter is the maximum eccentricity. The eccentricity of a node $v$ is the maximum distance from $v$ to all other nodes in $G$. If $G$ is disconnected, the eccentricity of a node $v$ is infinite. A diameter algorithm adapted based on the 

The Algorithm 3: the middle node algorithm is developed in order to find the middle point in a connected component of the induced acyclic subgraph. Since the input graph for finding the middle node is either a tree or a forest, we iteratively remove the nodes $x$ (and the adjunct edges) whose degrees satisfy $d(x) = 1$ or $d(x) = 0$. The last one node removed is the middle node, if the path is odd. One of the last two nodes removed is one of the two middle nodes, if the path is even. This middle node algorithm is implemented as below.

After the two steps of the node removal process, the induce subgraph satisfies the conditions as described in Theorem 1. We name the node $x \in V$ removed from $G$ as a **removed node**. The complementary induced subgraph $C_Ind_G(x)$ is called the induced subgraph at level node "$x$." All the removed nodes and the associated components are stored in a dictionary named **Subgraphs Dictionary (SD)** with removed nodes as keys and the associated components as values for recording this process.

**Algorithm 3:** The Middle Node Algorithm
**Input:**
The input graph, a tree or forest, is finite, undirected, and without loops or multiple edges. This input graph has at least ONE connected component whose diameter is greater than 2 edges.
**Step 1:**
Get a dictionary of the degrees of nodes in the input graph, namely "node_degree_dict", using the node name as keys and the degree value as values.
**Step 2:**
Find the keys which have values as 0 or 1, remove these nodes from the input graph to get the updated induced subgraph.
**Step 3:**

a. If the number of nodes in the updated induced subgraph is ZERO, the middle node is a node in the input graph (from step 1). Return this middle node.

b. If the number of nodes in the updated induced subgraph is not ZERO, clean the dictionary "node_degree_dict" and update the input graph with the updated induced subgraph. Then, start from step 1.

Algorithm 3: The Middle Node Algorithm

The number of removed nodes determines the number of iterations in both node removal and node adding processes so that we want to reduce the number of removed nodes to the greatest extend. By using the Algorithm 1, the basic cycles algorithm, we can break the cycles as many as possible at each removal so that we can reduce the graph to a tree with a minimum number of nodes removed. And by removing the middle node of the trees using the Algorithm 2, the diameter algorithm and Algorithm 3, the middle node algorithm, the diameter of the remaining trees are minimized, which is also minimizing the number of the node removed.

**Phase II: Adding Nodes and Conquering**

We consider a collection of problems that involve finding a feasible subset of the input of maximum weight. The input contains a collection of $n$ distinguished elements, each carrying an associated nonnegative rational weight. Each set of distinguished elements uniquely induces a candidate for a solution, which we assume is efficiently computable from the set. The weight of a solution is the sum of the weights of the distinguished elements in the solution.

Halldorsson defines such a partitioning structure as the hereditary property (Halldorsson, 2000). A property is said to be hereditary if whenever a set $S$ of distinguished element corresponds to a feasible solution, any subset of $S$ also corresponds to a feasible solution. A property is semi-hereditary if under the same circumstances, any subset $S'$ of $S$ uniquely induces a feasible solution, possibly corresponding to a superset of $S'$. Theorem 2 is based on this partitioning idea.

Let's take an example to explain Theorem 2. Given a weighted graph $G_3$ as Figure 3, the nodes, edges, node indexes, and weights associated is shown in Figure 3. Assuming node '3' is the removed node, according to Theorem 2, the Compare Set at level node '3' is the node set {'0', '3', '6'} circled in red in Figure 3, and the Preliminary Set at level node '3' is the node set {'0', '2', '5', '6'} circled in red in Figure 4. The $MWIS(G_3)$ is either the set {'0', '3', '6'} or {'0', '2', '5', '6'}. Since the set {'0', '2', '5', '6'} has a total weight 12 versus the total weight of {'0', '3', '6'}, which is 11, the $MWIS(G_3)$ is the set {'0', '2', '5', '6'}. In Figure 3, the induced subgraph in blue circles is the CSS, which is the $C_Neig_Ind_G(\{'3'\})$ plus node '3'. In Figure 4, the complementary induced subgraph, $C_Ind_G(n)$ in the green circle is the PSS at level node $n$.

**Theorem 2:** For Recurrence
For a given graph $G = (V, E)$, remove one node $n \in V$ (the removed node) to get the complementary induced subgraph $C_Ind_G(n)$. Let $MWIS(G)$ denote the MWIS of graph $G$ and let $AMIS(G)$ denote the AMIS of graph $G$.

(i) For the MWIS case, the $MWIS(G)$ is either the $MWIS[C_Ind_G(n)]$ or the maximum weighted independent set that has node $n$ as an element in graph $G$, $\{n\} \cup MWIS[C_Neig_Ind_G(n)]$. We name the $MWIS[C_Ind_G(n)]$as the **Preliminary Set** at level node $n$, and the $C_Ind_G(n)$ as the **Preliminary Set Subgraph (PSS)** at level node $n$. Similarly, we name the set $\{n\} \cup MWIS[C_Neig_Ind_G(n)]$ as the **Compare Set** at level node $n$, and the $C_Neig_Ind_G(n)$ with node $n$ as the **Compare Set Subgraph (CSS)** at level node $n$.

(ii) For the AMISL case, the AMIS of the complementary induced subgraph $C_Ind_G(n)$ is formally $AMIS[C_Ind_G(n)]$. All maximal independent sets which has node $n$ as an element in each of the all maximal independent sets in graph $G$ is formally $AMIS[Ind_G(n) \cup C_{Neig_{Ind\,G}}(n) \cup \{n\}]$. The all maximal independent set of $G$, $AMIS(G)$, is the union of the $AMIS[C_Ind_G(n)]$ and $AMIS[Ind_G(n) \cup C_Neig_Ind_G(n) \cup \{n\}]$. Note that if any maximal independent set in the AMISL outputs is a subset of another set in AMISL output sets in the union process. The subset is eliminated, since it is no longer a maximal independent set in the induced subgraph with node $n$. We name the $AMIS[C_Ind_G(n)]$ as the **AMISL Preliminary Sets** at level node $n$. Similarly, we name the $AMIS[Ind_G(n) \cup C_Neig_Ind_G(n) \cup \{n\}]$ as the **AMISL Compare Sets** at level node $n$.

**Proof of Theorem 2:** by contradiction
(Since the MWIS and AMISL algorithms follow the same structure, we only prove the MWIS case here.) As the conditions described in Theorem 2, assuming all three statements always hold:

1. The Preliminary Set is the MWIS of $C_Ind_G(n)$, $MWIS[C_Ind_G(n)]$;
2. The Compare Set is $\{n\} \cup MWIS[C_Neig_Ind_G(n)]$;
3. There exists an Assumption Set in $G$. The Assumption Set is a maximal independent set that has a total weight greater than that of either the Preliminary Set or the Compare Set.

In the same graph $G$, since the Assumption Set, a maximal independent set in $G$, has a total weight greater than the total weight of the Compare set, and the Compare Set has the maximum possible total weight of the maximal independent set has node $n$ as one element, the Assumption Set cannot contain node $n$ as an element. Because the Preliminary Set has the maximum possible total weight of the maximal independent set in the complementary induced subgraph $C_Ind_G(n)$, so that the maximum possible total weight of a maximal independent set without node $n$ as an element is equal to the total weight of the Preliminary Set. Since the Assumption Set cannot contain node $n$ as an element, then it must be a maximal independent set in the complementary induced subgraph $C_Ind_G(n)$ and its total weight is no greater than the total weight of the Preliminary Set. It is a Contradiction with statement 3, which implies that such an Assumption Set does not exist. ■

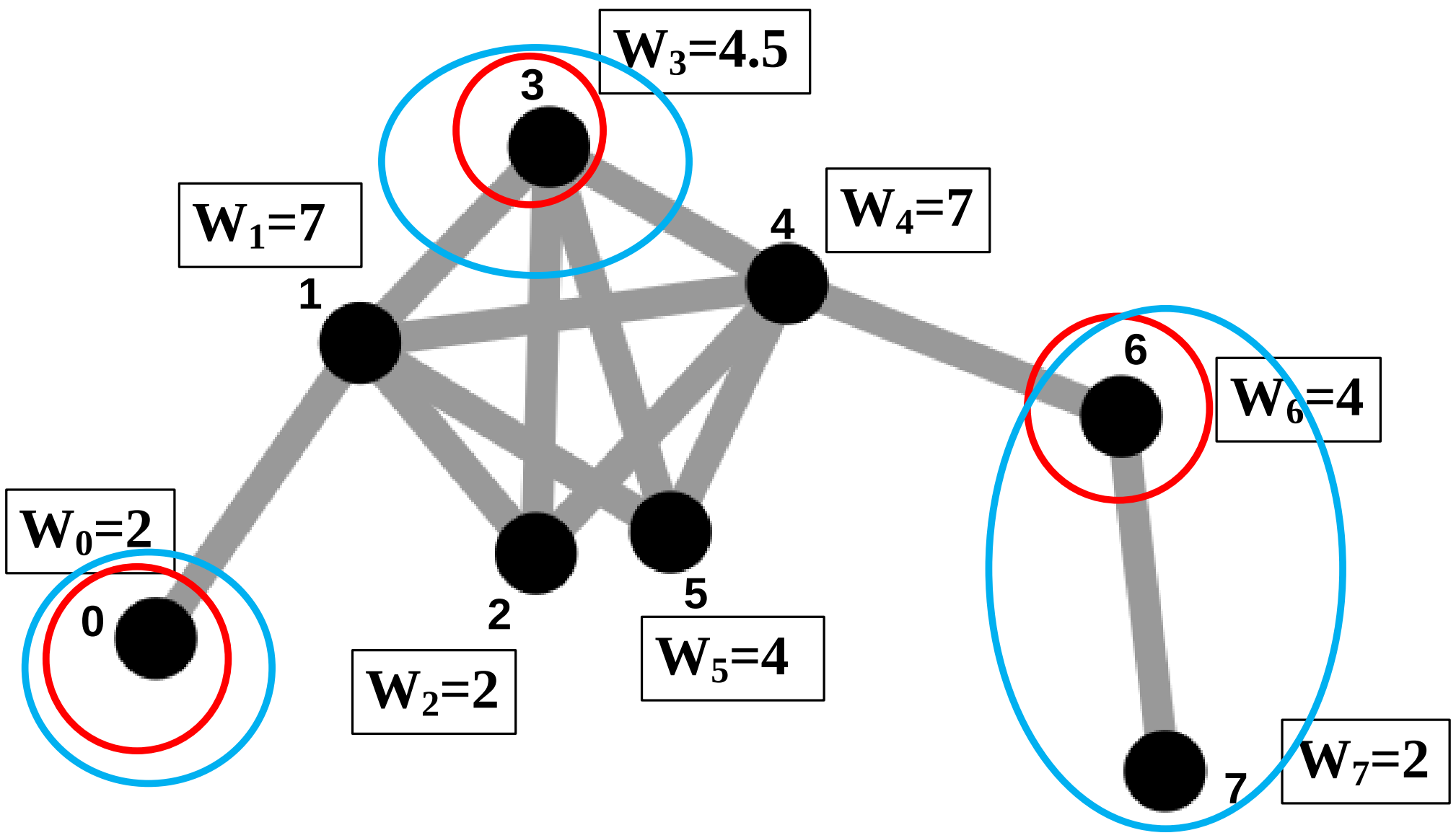

Figure 3. Compare Set at Level Node '3'

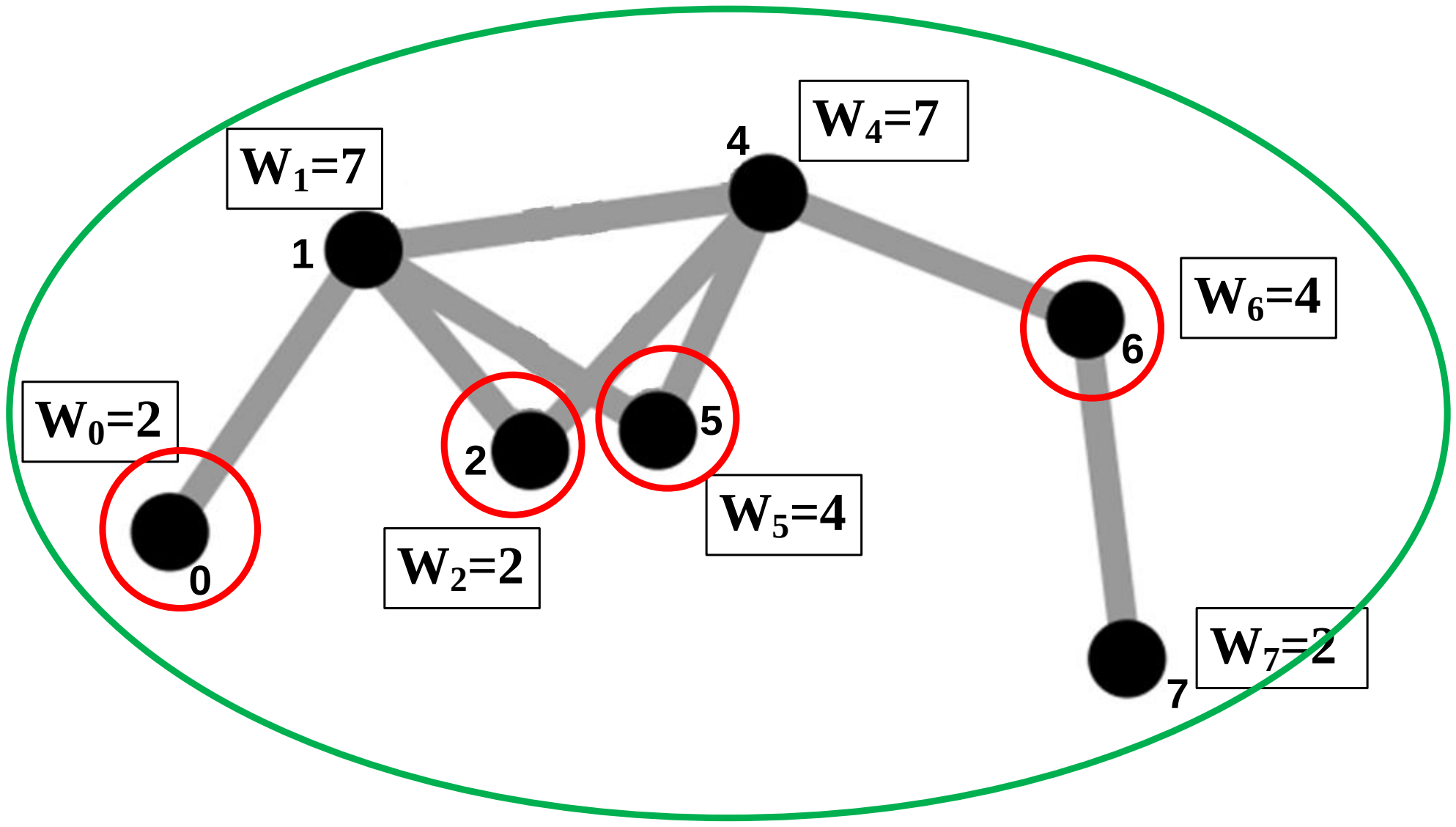

Figure 4. Preliminary Set at Level Node '3'

In order to further understand Theorem 2, suppose we decide to place a node $v$ into a given maximum weighted independent set. It then suffices to search only in the non-neighborhood of $v$, $C_Neig_Ind_G(v)$, for the remaining nodes in the set. This suggests a natural heuristic, the greedy method. We can specify its result formally as,

$$choose: v \in V$$

$$MWIS(G) \leftarrow \{'v'\} \cup MWIS[C_Neig_Ind_G(v)]$$

This rapid accumulation of an independent set by recursively looking at non-neighborhoods is attractive. Yet it remains disconcerting to completely ignore the neighborhoods of the pivot nodes, which may contain much larger weighted independent sets. Indeed, if we make a bad choice of a pivot node, we may be left with a minuscule set of independent vertices where there were plenty; thus, Greedy performs poorly in the worst case.

We are led to another rule for searching for an independent set. As before, choose a vertex and search in the non-neighborhood of that node. But this time also searches in the neighborhood of the pivot node, which makes the search area as $C_Ind_G(v)$, and use whichever result has a heavier total weight. More formally,

$$choose: v \in V$$

$$MWIS_AS(G) \leftarrow \max(\{'v'\} \cup MWIS_AS[C_Neig_Ind_G(v)], MWIS_AS[C_Ind_G(v)])$$

The discussions above are resulting in Algorithm 4, MWIS algorithm structure (MWIS_AS), as below:

```
MWIS_AS (G), G is a weight undirected graph.
Begin
        If G = ∅, then return [∅]
        Choose some v ∈ V
        [MWIS_1]← MWIS_AS[C_Ind_G(v)]
        [MWIS_2]← MWIS_AS[C_Neig_Ind_G(v) ∪ {'v'}]
        return (larger weight of (MWIS_1, MWIS_2))
End
```

Algorithm 4: MWIS Algorithm Structure

AMISL algorithm follows the same structure, but we need to define a particular function called the Special Union. Assuming $SS_1$ and $SS_2$ are two sets of sets, the Special Union, $Spec_ \cup (SS_1, SS_2)$, which is a set, which is the union of all the sets in $SS_1$ and $SS_2$, and no set in $Spec_ \cup (SS_1, SS_2)$ is a subset of another set. This is resulting in Algorithm 5, AMISL algorithm structure, (AMISL_AS) as below:

```
AMISL_AS (G), G is a weight undirected graph.
Begin
        If G = ∅, then return [∅]
        Choose some v ∈ V
        [AMIS_1]← AMISL_AS[C_Ind_G(v)]
        [AMIS_2]← AMISL_AS[C_Neig_Ind_G(v) ∪ {'v'}]
        return (Spec_∪(AMIS_1, AMIS_2))
End
```

Algorithm 5: AMISL Algorithm Structure

# 5 Construction of the Algorithms

From Theorem 1, we illustrate that the base cases for the divide and conquer algorithm structure. The base cases are constructed by removing nodes and the adjacent edges. We iteratively remove one node at a time by maximizing the number of cycles that the node belongs to in a cycle basis of the input graph or the current induced subgraph.

Subgraphs Dictionary (SD) is used to record this procedure. In SD, each node removed is the key and node sets of the connected components in the induced subgraphs as values of the keys, until the induced subgraphs satisfy the Theorem 1 conditions.

The node adding procedures that are illustrated in Figure 5, is based on Algorithm 4 and Algorithm 5. Assume there are $m$ removed nodes for computing the MWIS or AMIS of graph $G$, the CSS and the PSS denote the Compare Set Subgraph and the Preliminary Set Subgraph, respectively. The MWIS algorithm or the AMISL algorithm needs to be executed on the CSS at level node $l$, $l \in \{1,2, \dots, l, \dots, m\}$, with $n_l$ removed nodes to find the MWIS or the AMIS, respectively.

For the MWIS case, according to Theorem 2 and Algorithm 4, we can get the desired MWIS set by comparing the Compare Set and the Preliminary Set at each level of the removed node. The MWIS found at each level of the removed node is recorded in the subgraph MWIS dictionary (SMWISD): the current induced subgraph (the PSS plus the removed node at the level) is the key, and the MWIS found is the value. The SMWISD is used for searching the MWIS of the connected components, which is part of the Preliminary Set at the level.

For the AMISL case, according to Theorem 2 and Algorithm 5, we can get AMIS by comparing and merging the AMISL Compare Sets and the AMISL Preliminary Sets at each level of the removed node. The AMIS found at each level of the removed node is recorded in the subgraph AMIS dictionary (SAMISD): the current induced subgraph (the PSS plus the removed node at the level) is the key, and the AMIS found is the value. The SAMISD is used for searching the AMIS of the connected components, which is part of the AMISL Preliminary Set at the level.

Together with Corollary 1, recurrence can be set up by adding the removed nodes back to the graph in the reverse order from the CUSs till getting the whole original graph. At each level of the removed node, the Preliminary Set and the AMISL Preliminary Set can be found as follows. For the MWIS case, we can get the Preliminary Set by aggregating the MWIS of each connected component of the current induced subgraph (without the removed node) according to the key-value pair in the SD. These MWISs are found by searching the SMWISD or computed according to Theorem 1. For the AMISL case, following the Theorem 1 and Corollary 1, we can merge the AMISs of all connected components of the current induced subgraph according to the key-value pair in the SD to get the AMISL Preliminary Set. These AMISs are found by searching the SAMISD or computed according to Theorem 1.

While adding nodes back to get the Compare Set and the AMISL Compare Set, we follow the node adding heuristics for finding the Compare Set as below:

1. Get CSS, which is the induced subgraph by removing all neighbors of the removed node added; the removed node is included in the CSS.
2. Get the MWIS or AMIS of the CSS.
3. If the CSS getting from (1) does not satisfy the Theorem 1 conditions, perform the algorithm on this subgraph.

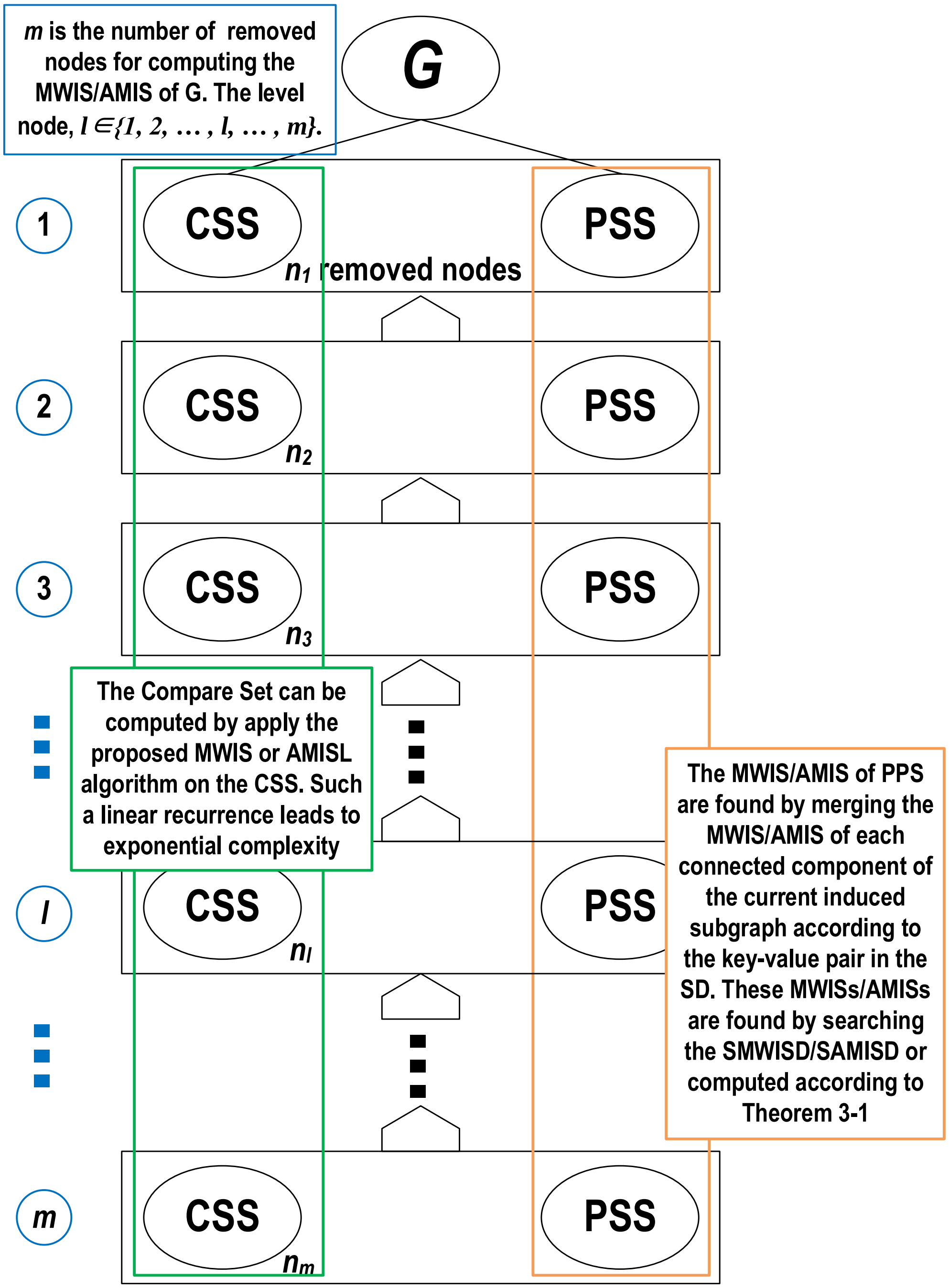

Figure 5. The Node Adding Procedures

Thus, the Algorithm A1 MWIS and Algorithm A2 AMISL can be constructed as below:

**Algorithm A1 MWIS**: A hybrid heuristic algorithm for MWIS problem
**Input**: a weighted graph $G$
**Output**: MWIS of graph $G$.
**Initializing**: subgraphs dictionary (SD) = {}; subgraph MWIS dictionary (SMWISD) = {}; 'last key' vertex = null.
**Begin:**

**(1.1)** **From step (1.1.1) to (1.1.5)** Based on the input graph, find and remove the nodes one at a time, based on the node removal procedures, and update the SD: each node removed is the key and vertices sets of the connected components in the induced subgraphs as values of the keys, until the induced subgraphs satisfy the Theorem 1 conditions.

**(1.1.1)** If the input graph satisfies the Theorem 1 conditions, go to step (1.2); if the input graph does not satisfy the Theorem 1 conditions, remove a vertex (the key in SD) and edges attached to it following the node removal steps in section 4.1, and get the component subgraphs vertices set(s) (value with the key);

**(1.1.2)** Update SD with the key-value pair;

**(1.1.3)** For each connected subgraph, exam whether it satisfies the Theorem 1 conditions;

**(1.1.4)** For those who do not satisfy Theorem 1 conditions, input these subgraphs to step (1.1.1); If the Theorem 1 conditions are satisfied, go to (1.1.5)

**(1.1.5)** When all subgraphs satisfy Theorem 1 conditions, return the latest SD and go to step (1.2).

**(1.2)** Get the Preliminary Set by aggregating the MWIS of each connected component of the induce subgraph according to the last key-value pair in SD. These MWISs are found by searching the SMWISD or computed according to Theorem 1.

**(1.3)** If 'last key' vertex = null, Compare Set is $\emptyset$; if not add the 'last key' vertex to the induced subgraph from (1.2) and follow the node adding heuristics to find the Compare Set at the level 'last key'.

**(1.4)** Get the set with maximum total weight among the two sets: Preliminary Set and Compare Set at the level 'last key'. This set is the MWIS at the level 'last key' (the MWIS of the induced subgraph of the last level in SD). Update the SMWISD: the current induced subgraph from (1.3) is the key, and the MWIS found is the value.

**(1.5)** Update SD by removing the last key-value pair. If the updated $SD = \{\}$, return the MWIS from step (1.4); if not, go to step (1.2).

Algorithm A1 MWIS: A Hybrid Heuristic Algorithm for MWIS Problem

For better describing the algorithms we proposed in this section, we provide a walk-through of Algorithm A1 as well as all the terms in detail with a simple example in Appendix I. In the following section, we discuss the complexity of the proposed algorithms, and the means to improve the computational speed.

**Algorithm A2 AMISL**: A hybrid heuristic algorithm for AMISL problem
**Input**: a weighted graph $G$
**Output**: MWIS of graph $G$.
**Initializing**: subgraphs dictionary (SD) = {}; subgraph AMIS dictionary (SAMISD) = {}; 'last key' vertex = null.
**Begin:**

**(2.1)** **From step (2.1.1) to (2.1.5)** Based on the input graph, find and remove the vertices one at a time, based on the vertices removal procedures, and update the SD: each vertex removed is the key and vertices sets of the connected components in the induced subgraphs as values of the keys, until the induced subgraphs satisfy the Theorem 1 conditions.

**(2.1.1)** If the input graph satisfies the Theorem 1 conditions, go to step (2.2); if the input graph does not satisfy the Theorem 1 conditions, remove a vertex (the key in SD) and edges attached to it following the node removal steps in section 4.1, and get the component subgraphs vertices set(s) (value with the key);

**(2.1.2)** Update SD with the key-value pair;

**(2.1.3)** For each connected subgraph, exam whether it satisfies the Theorem 1 conditions;

**(2.1.4)** For those who do not satisfy Theorem 1 conditions, input these subgraphs to step (2.1.1); If the Theorem 1 conditions are satisfied, go to (2.1.5)

**(2.1.5)** When all subgraphs satisfy Theorem 1 conditions, return the latest SD and go to step (2.2).

**(2.2)** Following the Theorem 1 and Corollary 1, merge the AMISs of all connected components of the induce subgraph according to the last key-value pair in SD to get the AMISL Preliminary Set. These AMISs are found by searching the SAMISD or computed according to Theorem 1.

**(2.3)** If 'last key' vertex = null, Compare Set is $\emptyset$; if not add the 'last key' node to the induced subgraph from (2.2) and follow the node adding heuristics to find AMISL Compare Sets at the level 'last key'.

**(2.4)** Get the Special Union of the two sets of sets: AMISL Preliminary Set and AMISL Compare Set at the level 'last key'. Note that if any maximal independent set in the union is a subset of another set in this union process, eliminate this set from the union. This union is the AMISL output at the level 'last key' (the AMIS set of the induced subgraph of the last level in SD). Update the SAMISD: the current induced subgraph from (2.3) is the key, and the AMIS found is the value.

**(2.5)** Update SD by removing the last key-value pair. If the updated $SD = \{\}$, return the AMIS from step (2.4); if not, go to step (2.2).

**(2.6)** Find the MWIS based on the AMIS.

Algorithm A2 AMISL: A Hybrid Heuristic Algorithm for MWIS/AMISL Problem

# 6 Reducing the Complexity of the Algorithm Using Approximation Algorithms

## 6.1 Discussion on the Complexity

The runtime of the proposed Algorithm A1 and A2 highly depends on the input graph. In the Algorithm A1, the node adding procedures through step (1.2) to step (1.5), the Preliminary Sets are computed based on the CUS, or they may inherit the MWIS of previous induced subgraph before adding the node. By searching the dictionary, which stores the results of previous node adding steps, computations for Preliminary Sets are at a low cost. But computations for Compare Sets may require executing Algorithm A1 on the CSSs according to the node adding heuristics. This leads to exponential complexity.

Let us take the graph $G_6$ in Figure 6 as an example to illustrate the complexity of the proposed algorithm structure, to simplify the problem, assuming weights of the vertices are the same as the vertex index.

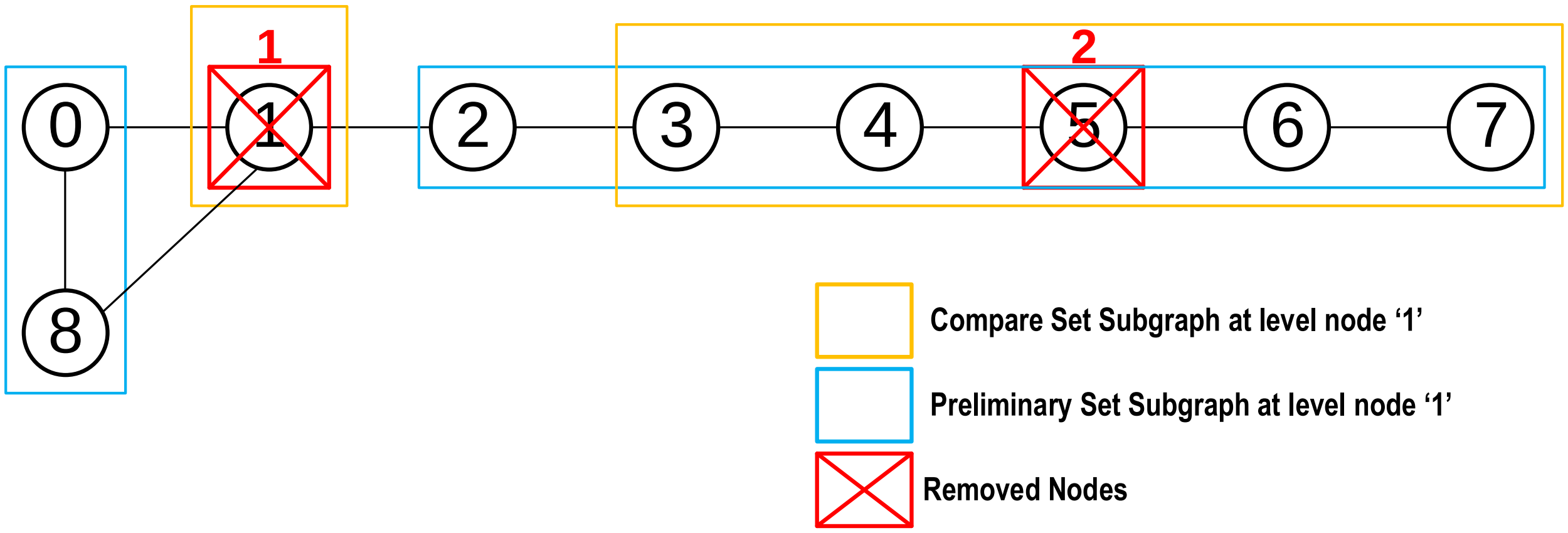

Figure 6. A sample graph with 9 vertices

Based on step (1.1) in Algorithm A1,

$$SD = \{'1': [\{'0','8'\},\{'2','3','4','5','6','7'\}],'5': [\{'2','3','4'\},\{'6','7'\}]\}$$

At level node '5', the Preliminary Set is {'2','4','7'} and the Compare Set is {'3','5','7'} in the subgraph induced by nodes, $\{'5','2','3','4','6','7'\}$. The MWIS as level node '5' is {'3','5','7'}. At level node '1', based on the step (1.4), the Preliminary Set is the union the two MWIS of the two induced subgraphs (in the blue boxes), $Ind_{G_6}(\{'0','8'\})$ and $Ind_{G_6}(\{'2','3','4','5','6','7'\})$. The MWIS of $Ind_{G_6}(\{'0','8'\})$ is simple to know. The MWIS of $Ind_{G_5}(\{'2','3','4','5','6','7'\})$ is the same as the MWIS at level node '5', which is {'3', '5', '7'}. But for the Compare Set, whenever the CSS does not satisfy the Theorem 2 conditions, we need to execute the Algorithm A1. Just like the CSS in the yellow boxes shown as Figure 7, it requires to execute Algorithm A1 to get the Compare Set at level node '1', which is {'1', '3', '5', '7'}. Such a linear recurrence leads to exponential complexity (Erickson, 2018). Note that, since Algorithm A2 follows a similar structure, but it is returning the AMIS at each step, the Algorithm A1 and A2 have the same complexity with the same input graph.

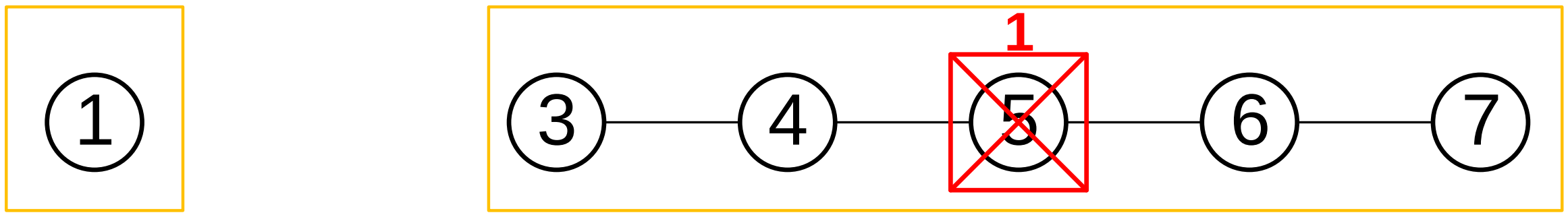

Figure 7. The CSS at Level Node '1'

## 6.2 Merging Approximation Algorithms with the Proposed MWIS Algorithm

Since calculations for the Compare Set slow down the execution of the proposed Algorithm A1 for the MWIS problem, we can speed up the computation by replacing Algorithm A1 on computing MWIS for Compare Sets with fast MWIS approximation algorithms. To illustrate this idea, we utilize two low complexity approximation algorithms to compute the Compare Set. Sakai et al. (Sakai et al., 2003) discuss greedy algorithms for the MWIS problem (GMIN-type algorithms). Two algorithms are the GMWIN and GMWIN2, which select a node of maximizing a node selection function, then remove it and its neighbors from the graph, and iterates this process on the remaining graph (induced subgraph) until no vertex remains. The set of selected nodes is the desired independent set. Let $G = (V, E, W)$ be a simple undirected graph with node set $V$, a set of edges $E$, and $W$ is a set of weight factors associated with element in $V$. Let $u, v \in V$, for each $v_i \in V$ $(0 \leq i \leq |I| - 1)$, the two node-selecting functions are:

(1) GWMIN: maximizing $\frac{W_u}{d_{G_i}(u)+1}$

(2) GWMIN2: maximizing $\frac{W_u}{\sum_{u \in Neig^+_{G_i}(u)} W_u}$

Where, $G_i$ is the remaining graph. We refer to the two simple greedy algorithms as Algorithm A3 GMWIN and Algorithm A6 GMWIN2, which are using the GWMIN and GWMIN2 node selection functions, respectively.

Let us consider the following framework of GMIN-type algorithms.

**Algorithm A3 GMWIN and Algorithm A6 GMWIN2,** GMIN-type Algorithm Framework
**INPUT**: A weighted graph G
**OUTPUT**: A maximal independent set in G
**begin**
    $I := \emptyset$; $i := 0$; $G_i := G$;
    **while** $V(G_i) \neq \emptyset$ **do**
        Choose a node based on a node-selecting function, say $v_i$, in $G_i$;
        $I := I \cup \{v_i\}$; $G_i + 1 := G_i[V(G_i) - Neig(v_i) + G_i(v_i)]$;
        $i := i + 1$;
    **od**
    Output $I$;
**end.**

Algorithm A3 and A6. The Algorithm GWMIN and Algorithm GMWIN2

As approximation algorithms, we are interested to know the lower bound of their accuracy. Sakai et al. (Sakai et al., 2003) proved Theorem 3 and Theorem 4 as the lower bounds of the accuracy of the two algorithms.

**Theorem 3.** Algorithm A3 GWMIN outputs an independent set of weight at least $\sum_{v \in V} \frac{W_v}{d_G(v)+1}$.

**Proof of Theorem 3:**

$$\sum_{i=0}^{|I|-1} W_{v_i} \geq \sum_{i=0}^{|I|-1}\left(\sum_{u\in Neig_{G_i}^{+}(v_i)} \frac{W_u}{d_{G_i}(u)+1}\right)$$
$$\geq \sum_{i=0}^{|I|-1}\left(\sum_{u\in Neig_{G_i}^{+}(v_i)} \frac{W_u}{d_G(u)+1}\right)$$
$$= \sum_{v\in V} \frac{W_v}{d_G(v)+1} \blacksquare$$

**Theorem 4.** Algorithm A6 GWMIN2 outputs an independent set of weight at least $\sum_{v\in V} \frac{W_v^2}{\sum_{u\in Neig_G^{+}(v)} W_u}$.

**Proof of Theorem 4:**
Let $I = \{v_1, v_2, \ldots, v_t\}$ be the independent set obtained by the algorithm. Let $f_G(v) = W_v / \sum_{u\in Neig_G^{+}(v)} W_u$.

$$\sum_{i=1}^{t} W_{v_i} \geq \sum_{i=1}^{t}\left(f_{G_i}(v_i) \times \sum_{u\in Neig_{G_i}^{+}(v_i)} W_u\right)$$
$$\geq \sum_{i=1}^{t}\left(\sum_{u\in Neig_{G_i}^{+}(v_i)} f_{G_i}(v_i) W_u\right) \quad \text{(from } f_{G_i}(v_i) \geq f_{G_i}(u)\forall u \in V(G_i))$$
$$\geq \sum_{v\in V(G)} f_G(v) W_v \quad \text{(from } f_{G_i}(u) \geq f_G(u)\forall u \in V(G))$$
$$= \sum_{v\in V} \frac{W_v^2}{\sum_{u\in Neig_G^{+}(v)} W_u} \blacksquare$$

With the approximation algorithms ready, we employ two different methods to merge an approximation algorithm with the proposed MWIS algorithm structure. Shown as Figure 8, in the step (1.3) of Algorithm A1, we denote the whole induced subgraph $G_l$ at the level node '$l$,' which is the PPS at the level node '$l$' plus node '$l$' (with the attached edges) in the node adding processes. Based on this assumption, the CSS at the level node '$l$' is the induced subgraph of $C_Neig_Ind_{G_l}(l)$ plus the node '$l$,' $C_Neig_Ind_{G_l}(l) \cup \{'l'\}$; the PPS is the complementary induced subgraph $C_Ind_{G_l}(l)$. We can either apply an approximation algorithm on the whole induced subgraph $G_l$ or the $C_Neig_Ind_{G_l}(l)$ for computing an MWIS as the Compare Set at the level node '$l$.' Formally, for the two approximation algorithms, GWMIN and GWMIN2, four merged MWIS approximation algorithms are as follows:

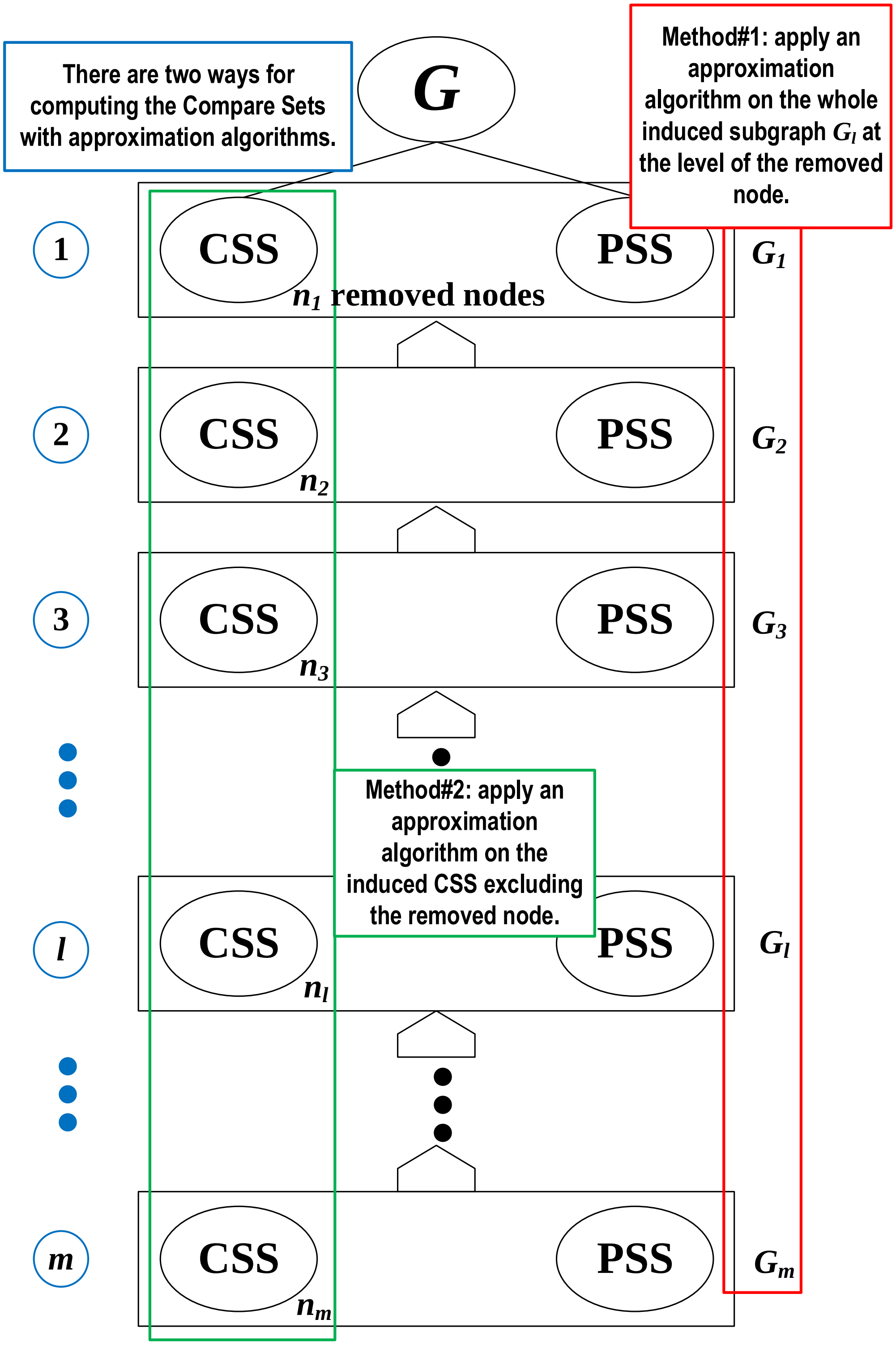

Figure 8. Merging Approximation Algorithms with the MWIS Algorithm Structure

(1) Algorithm A4 MWIS_CS_GWMIN: In the step (1.3) of Algorithm A1, when the CSSs do not satisfy the Theorem 1 conditions, instead of executing the Algorithm A1 on the CSSs, we compute Compare Sets based on the whole subgraph $G_l$ using Algorithm A3 GWMIN.

(2) Algorithm A5 MWIS_SubCS_GWMIN: In the step (1.3) of Algorithm A1, when the CSSs do not satisfy the Theorem 1 conditions, we use Algorithm A3 GWMIN to compute MWISs on the $C_Neig_Ind_{G_l}(l)$, then plus node '$l$' for Compare Set computations.

(3) Algorithm A7 MWIS_CS_GWMIN2: In the step (1.3) of Algorithm A1, when the CSSs do not satisfy the Theorem 1 conditions, we compute Compare Sets based on the whole subgraph $G_l$ using Algorithm A6 GWMIN2.

(4) Algorithm A8 MWIS_SubCS_GWMIN2: In the step (1.3) of Algorithm A1, when the CSSs do not satisfy the Theorem 1 conditions, we use Algorithm A3 GWMIN2 to compute MWISs on the $C_Neig_Ind_{G_l}(l)$, then plus node '$l$' for Compare Set computations.

According to Theorem 2, both composed MWIS approximation algorithms generate results no worse than the lower bound of the original approximation algorithms. In Algorithm A5 and Algorithm A8, the approximation algorithms are used on the $C_Neig_Ind_{G_l}(l)$, compare to Algorithm A4 and Algorithm A7, which are the approximation algorithms using the Algorithm A3 GWMIN and Algorithm A6 GWMIN2 on the whole subgraph $G_l$, respectively. By definition, the complementary neighbor induced subgraph, $C_Neig_Ind_{G_l}(l)$, is smaller than the whole induced subgraph $G_l$, because the node $n$ and its neighbors are not included in $C_Neig_Ind_{G_l}(l)$. Theoretically, the Algorithm A5 and Algorithm A8 should have better accuracy than the Algorithm A4 and Algorithm A7, respectively. And the Algorithm A5 and Algorithm A8 should have a faster computational speed than the Algorithm A4 and Algorithm A7, respectively. The computational experiments in the following section also justify these conjectures.

# 7 Computational Experiment on MWIS Algorithms

According to the proposed approach for the PPS problem discussed in our work (Sun et al., preprint), conflicting weighted graphs (general graphs) are created to test the scalability and accuracy of the algorithms in solving the PPS problem. Forty-three conflicting weighted graphs are created based on randomized PPS problems, from 5 nodes and 6 edges to 161 nodes and 4718 edges. The scalability analysis shows how the algorithms behave on the test graphs. It can be evaluated based on the computation time versus the different sizes of the test graphs, which measures by the node numbers and edge numbers of the different conflicting graphs. The accuracy refers to how likely the proposed approach can get to the optimum solution, MWIS. It can be measured by the average and the maximum error rate of all the test instances. The details of the results are shown in Appendix II. Note that all computational experiments in this work are performed on a virtual server at Syracuse University. The CPU is Intel Xeon E5-2699 with a fixed maximum speed at 2.3 GHz, and the memory is 32 GB. All the implementations mentioned in this work are in single threading.

Before we start the discussion on the scalability and accuracy, let us formally summarize all the MWIS algorithms to be tested as below:

- Algorithm A1 MWIS: the proposed exact MWIS algorithm.
- Algorithm A2 AMISL: the proposed exact AMISL-based MWIS algorithm.
- Algorithm A3 GWMIN: the GWMIN approximation algorithm from literature.
- Algorithm A4 MWIS_CS_GWMIN: it is an algorithm composed of Algorithm A1 and Algorithm A3. This algorithm computes Compare Sets based on the whole induced subgraph at each level using Algorithm A3 GWMIN.
- Algorithm A5 MWIS_SubCS_GWMIN: it is an algorithm composed of Algorithm A1 and Algorithm A3. This algorithm computes Compare Sets based on the induced CSSs, excluding the current removed node, using Algorithm A3 GWMIN.
- Algorithm A6 GWMIN2: the GWMIN2 approximation algorithm from literature.
- Algorithm A7 MWIS_CS_GWMIN2: it is an algorithm composed of Algorithm A1 and Algorithm A6. This algorithm computes Compare Sets based on the whole induced subgraph at each level using Algorithm A6 GWMIN2.
- Algorithm A8 MWIS_SubCS_GWMIN2: it is an algorithm composed of Algorithm A1 and Algorithm A6. This algorithm computes Compare Sets based on the induced CSSs, excluding the current removed node, using Algorithm A6 GWMIN2.

The computation time of Algorithms A1 and A2 changing with node number and edge number is shown in Figure 9 and Figure 10, respectively. Algorithms A1 and A2, as discussed in Section 6, can be exponentially slow on certain graphs. The computation time can be hours when there are about 140 nodes and 4000 edges. Although the worst case of the two algorithms can be exponentially slow, the application scenarios of the PPS problem considered here may not always be the worst case. Algorithms A1 and A2 match higher-order (order 4 or higher) polynomial trendlines, but they are faster than the exponential trendline.

Figure 11 and Figure 12 show how the computation time changing with node number and edge number on Algorithms A3 and A6, respectively. Algorithms A3 and A6 are the approximation algorithms from literature, and they are the fastest among the 8 algorithms. The computation time is less than one second on the test graphs. Algorithms A3 and A6 are in lower-order polynomial complexity on the test graphs. The difference in the complexity of the two algorithms is due to the different greedy functions of the two algorithms.

Figure 13 and Figure 14 show how the computation time is changing with node number and edge number on Algorithms A4, A5, A7, and A8, respectively. Algorithms A4, A5, A7, and A8 are the composed algorithms based on Algorithm A1 structure with MWIS approximation algorithms. They are slower than the approximation algorithms utilized, but they are still much faster than the exact MWIS algorithms. The computation time is less than 45 seconds on the test graphs. Algorithm A5 and A8 are faster than Algorithm A4 and A7, respectively. This result of computational experiments matches the conjectures in Section 6 that is the Compare Set computation is based on a smaller subgraph. And the Algorithm A7 and A8 are faster than Algorithm A4 and A5, respectively. This result also justifies that Algorithms A6 is faster than Algorithms A3 when the graph is relatively small (less than 3500 edges and

less than 135 nodes.)

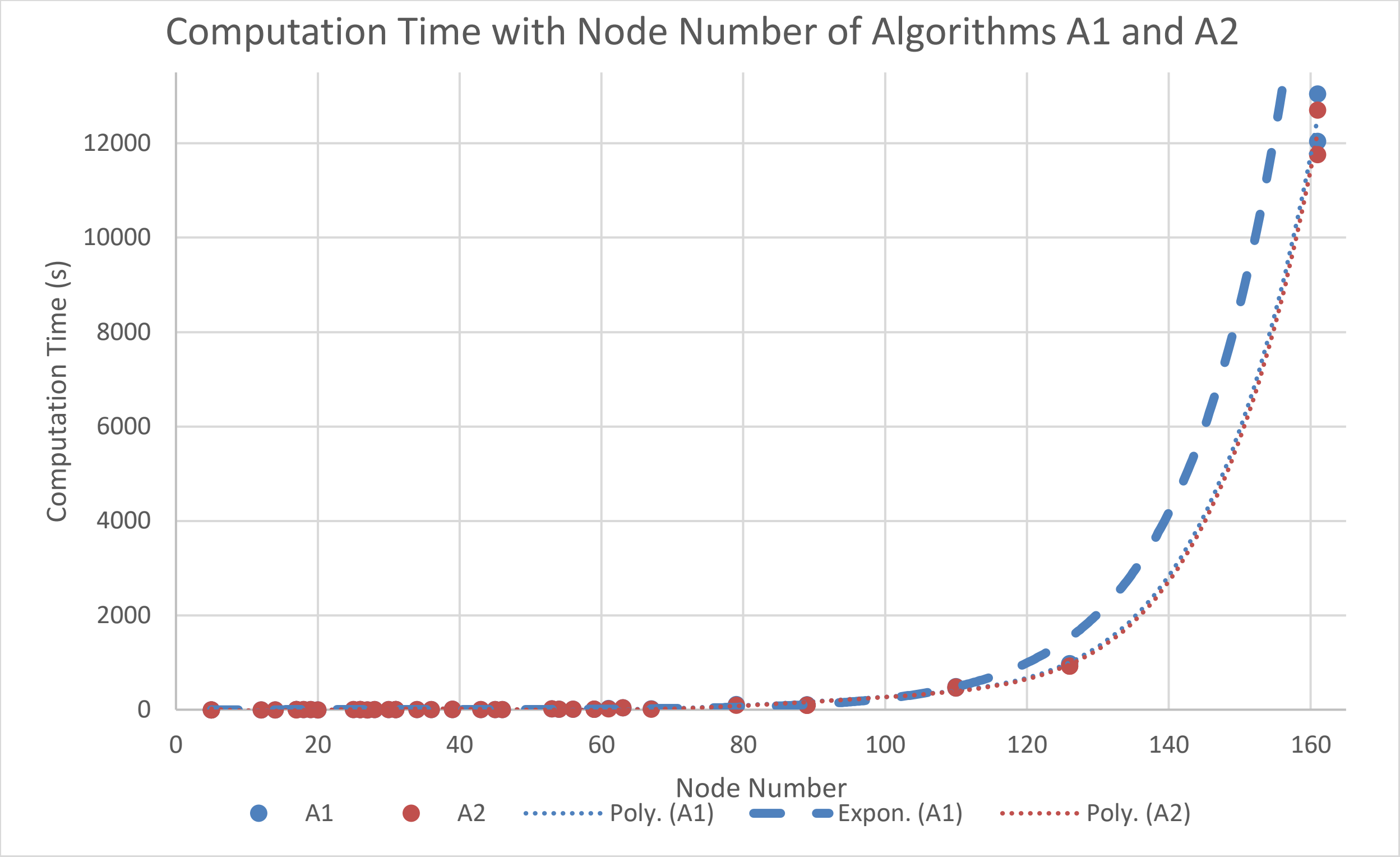

Figure 9. Computation Time with Node Number of Algorithms A1 and A2

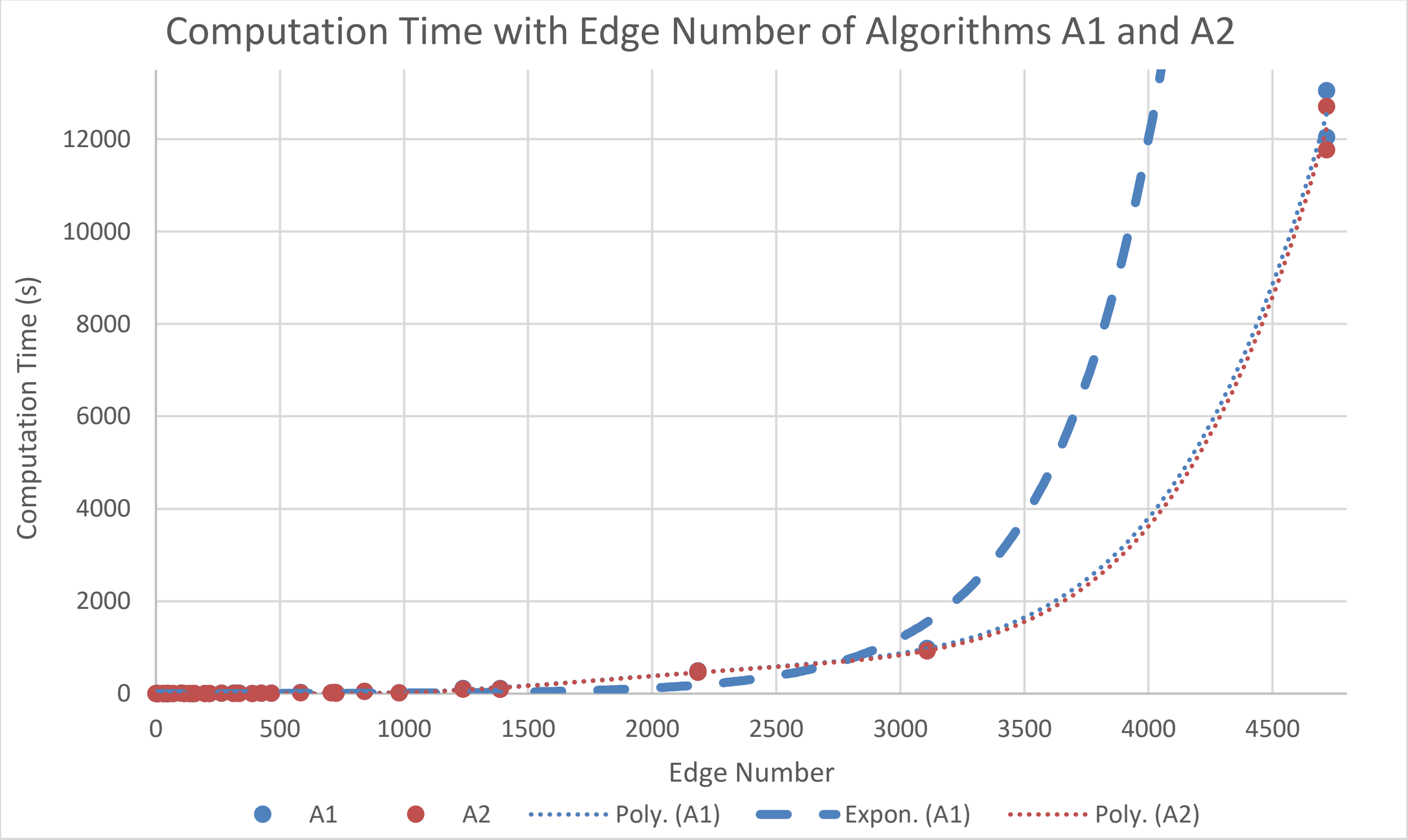

Figure 10. Computation Time with Edge Number of Algorithms A1 and A2

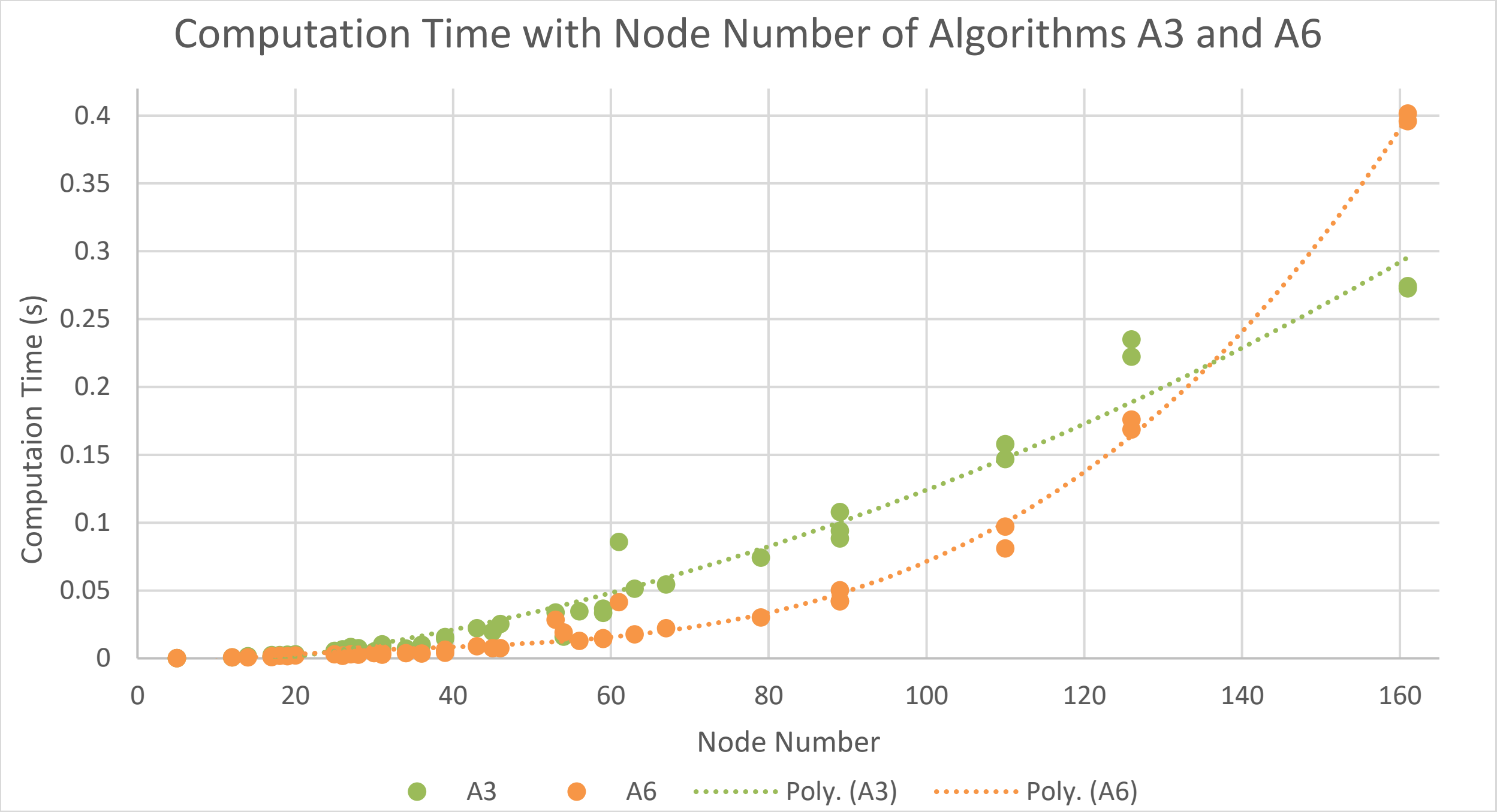

Figure 11. Computation Time with Node Number of Algorithms A3 and A6

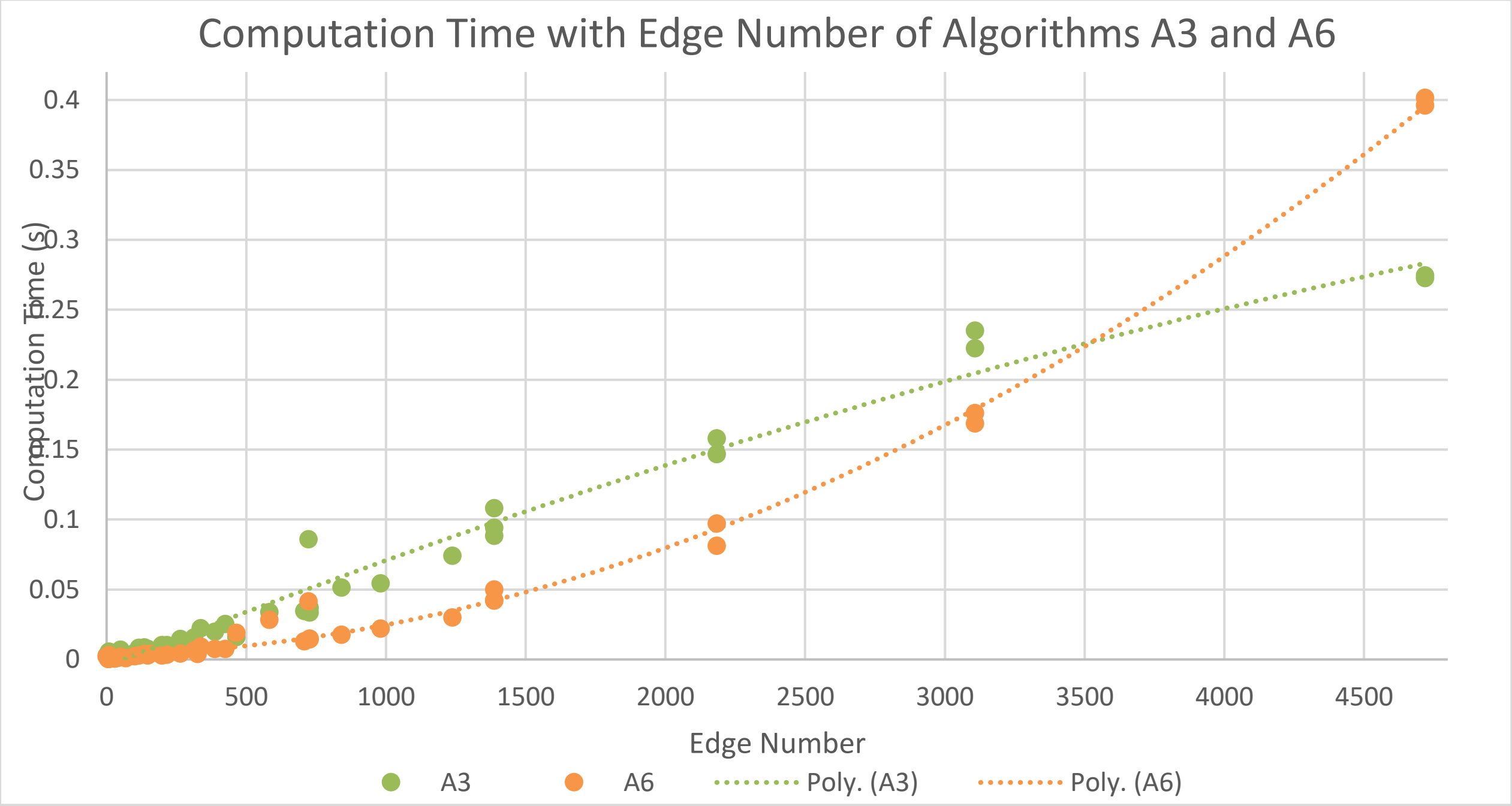

Figure 12. Computation Time with Edge Number of Algorithms A3 and A6

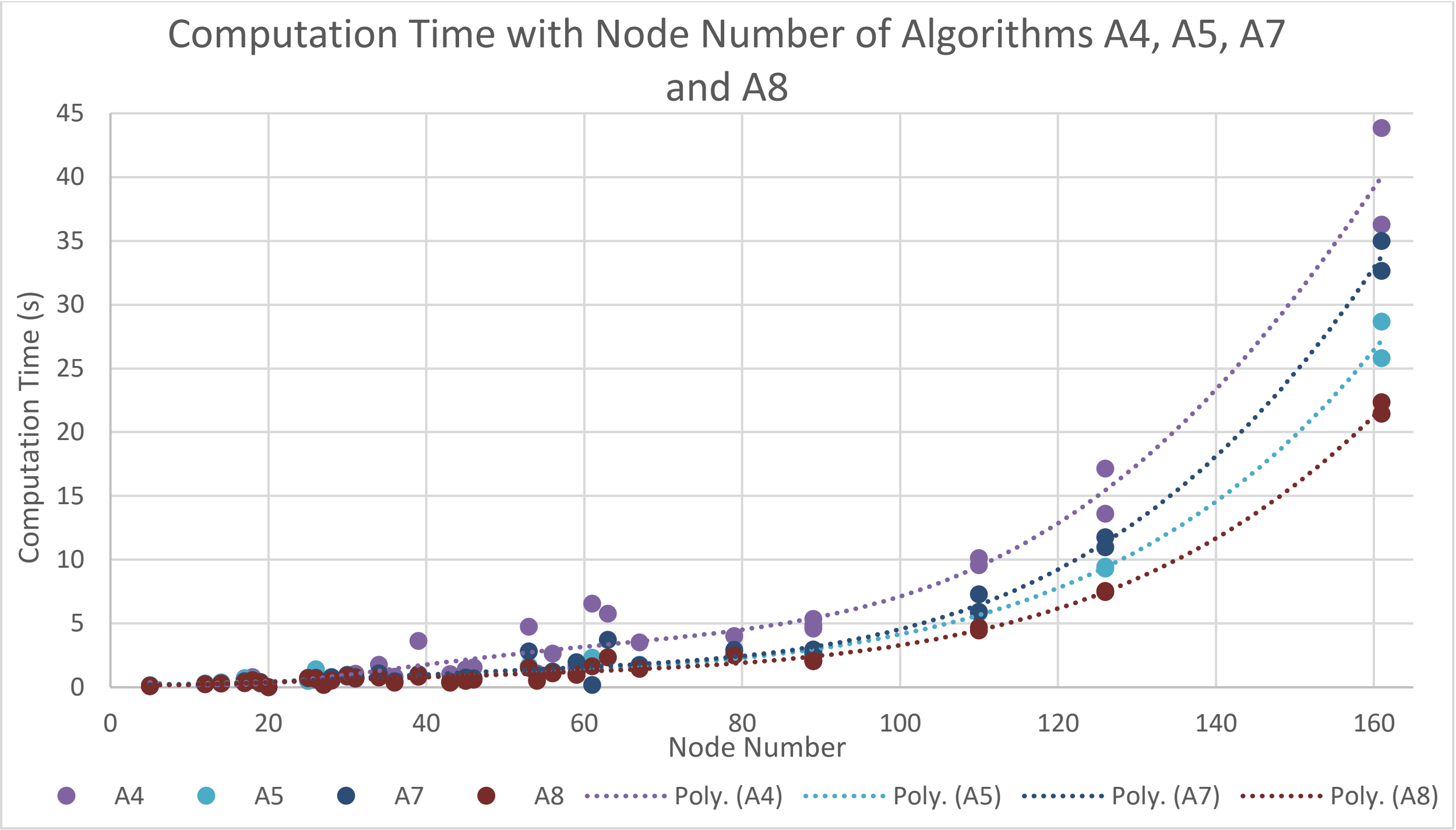

Figure 13a. Computation Time with Node Number of Algorithms A4, A5, A7 and A8

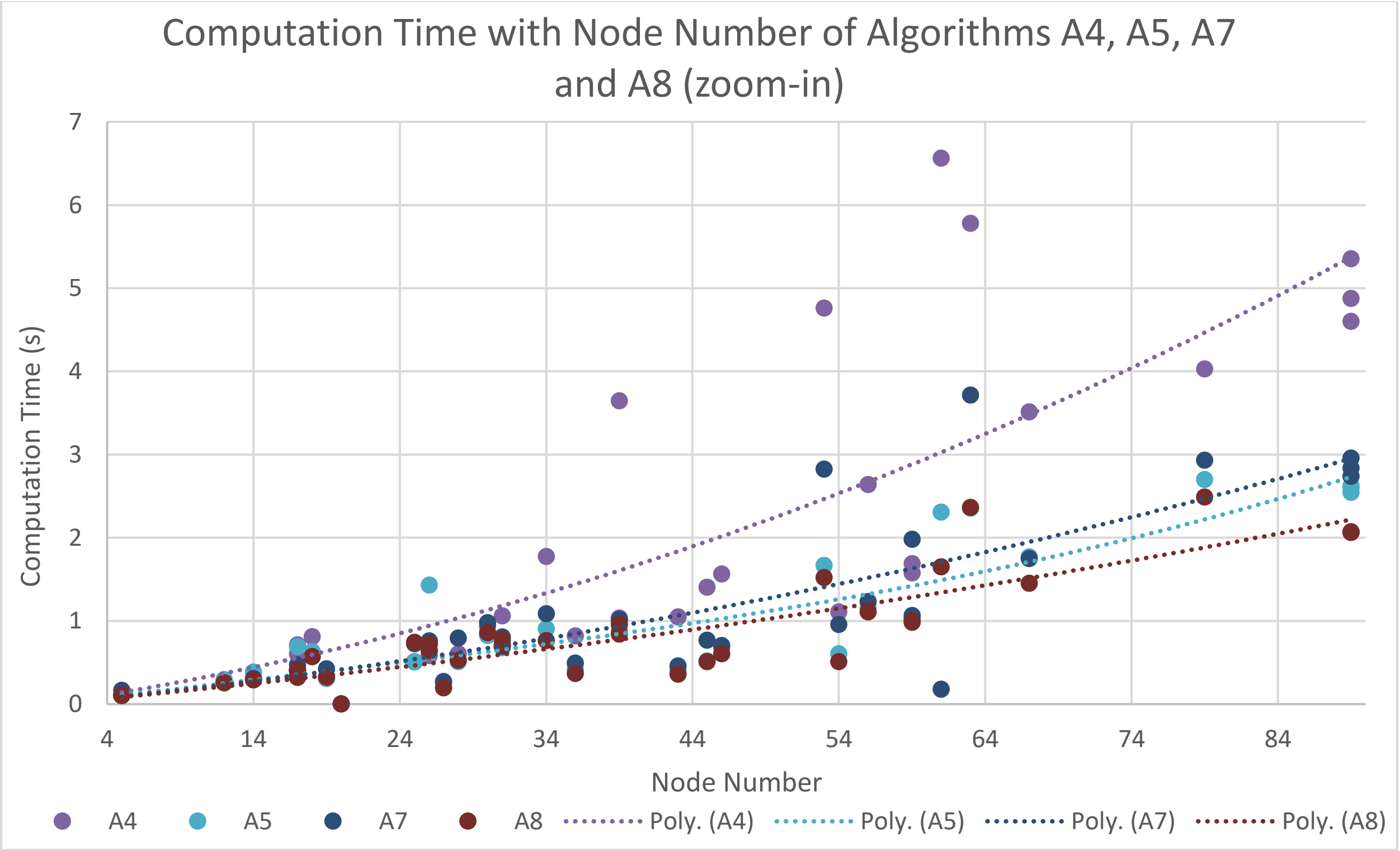

Figure 13b. Computation Time with Node Number of Algorithms A4, A5, A7 and A8 (zoom-in)

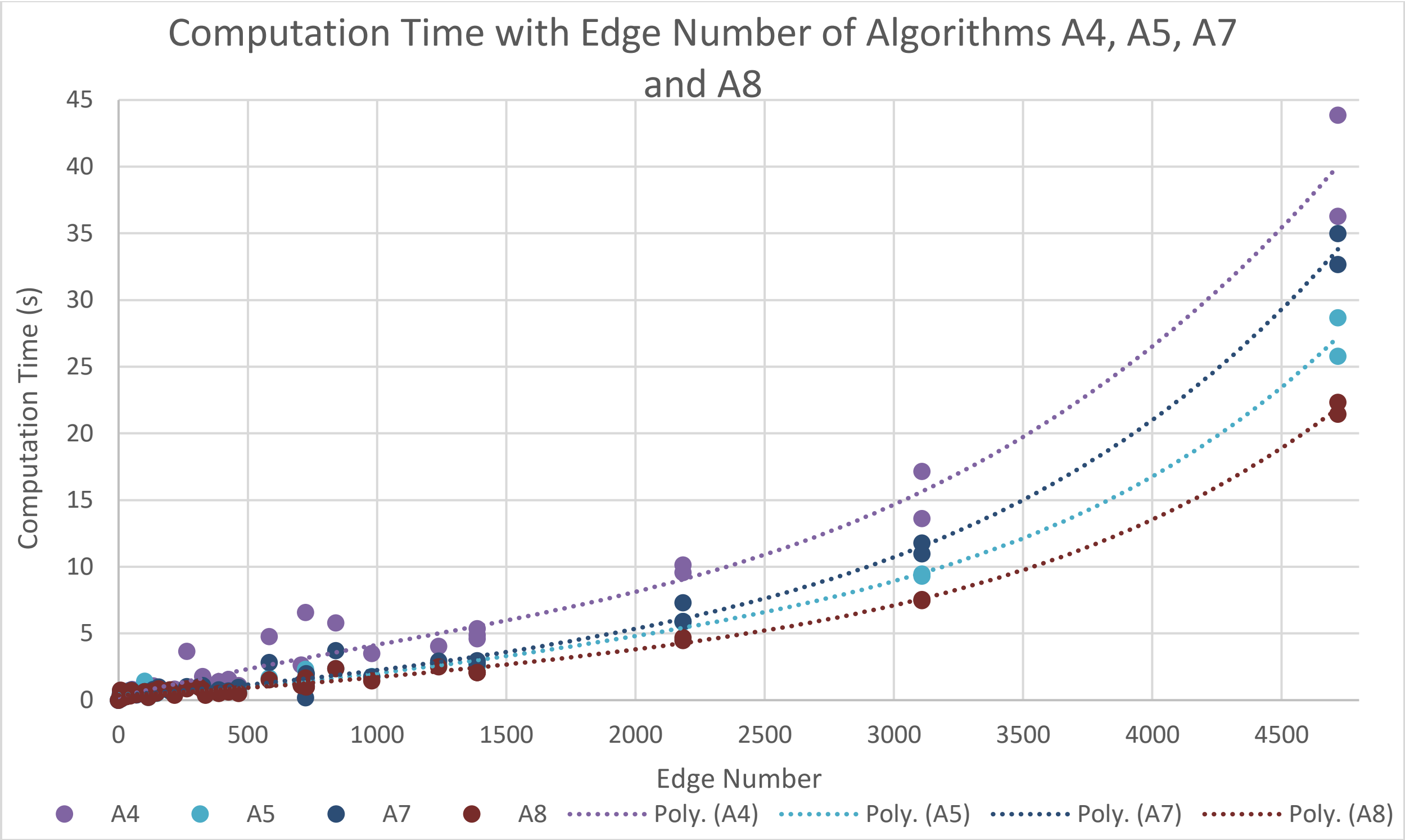

Figure 14a. Computation Time with Edge Number of Algorithms A4, A5, A7 and A8

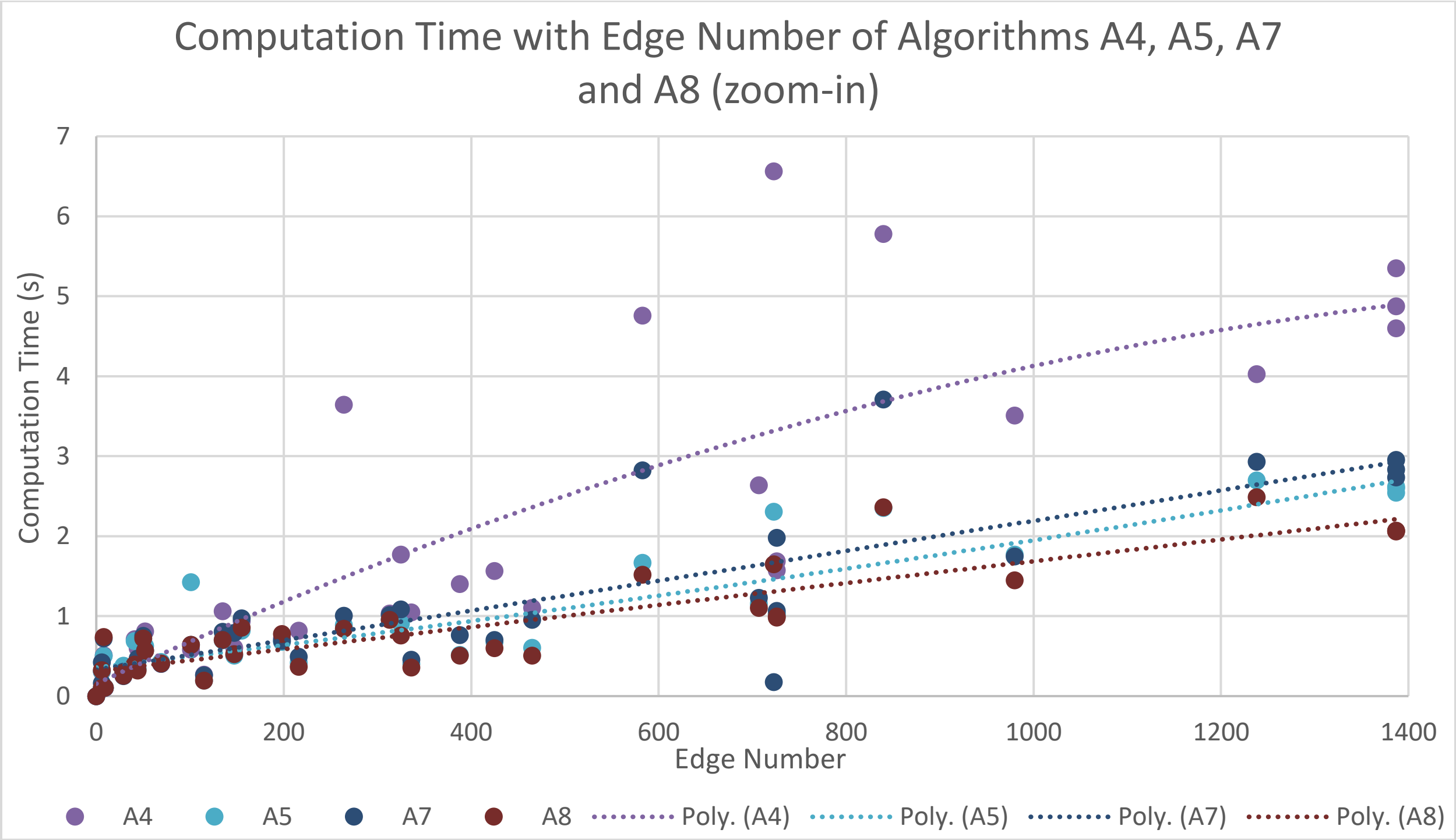

Figure 14b. Computation Time with Edge Number of Algorithms A4, A5, A7 and A8 (zoom-in)

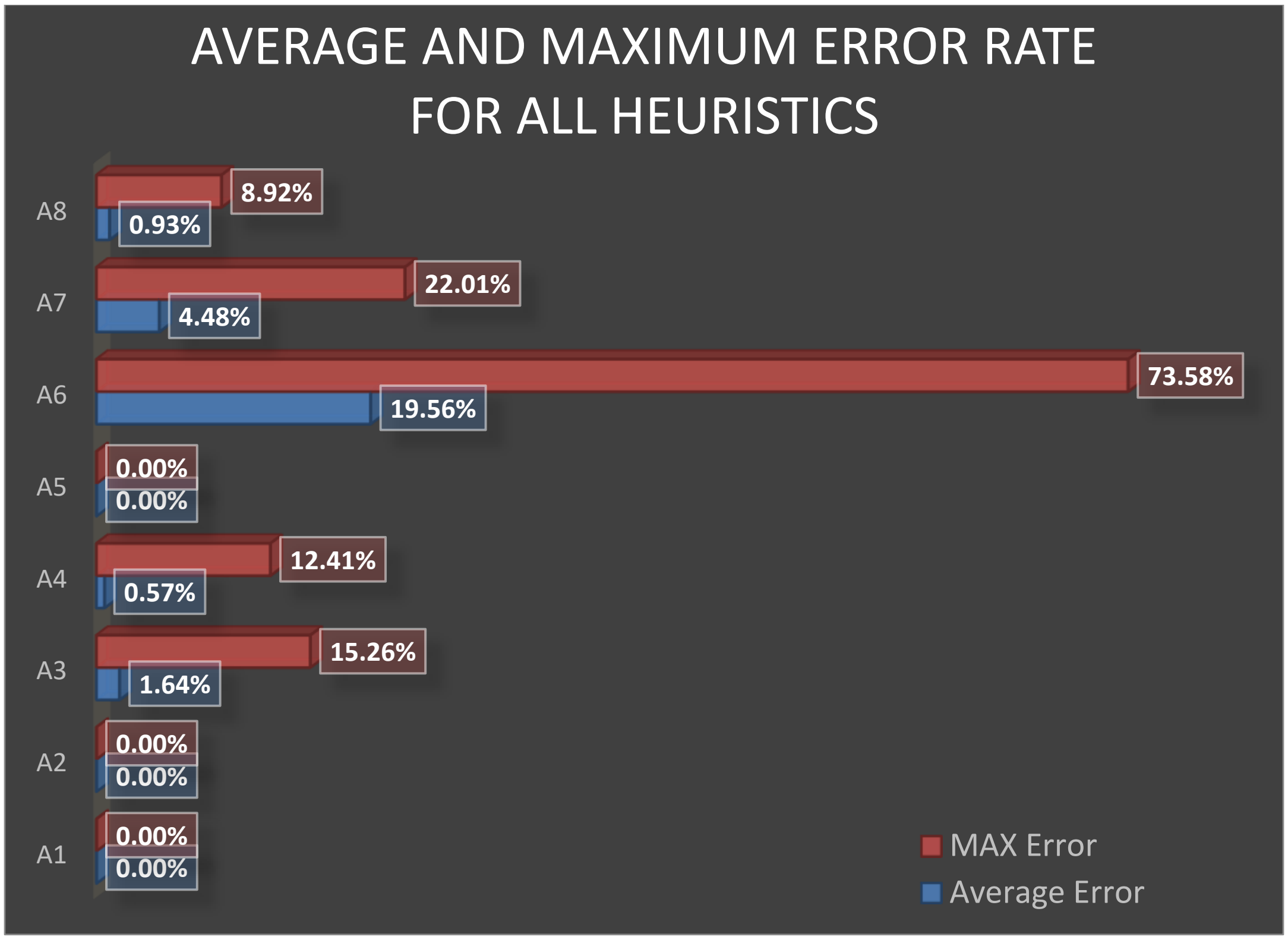

Figure 15. The Average and Maximum Error Rate for All Algorithms

Figure 15 shows the average and maximum error rate of the algorithms. Assume $W_{optimum}$ is the total weight of the optimum solution of the MWIS problem on the test graph, and $W$ is the total weight of the MWIS set found by the algorithm. The weight error rate is calculated using the function below.

$$Weight\ Error\ Rate = \frac{W - W_{optimum}}{W_{optimum}} \times 100\%$$

Note that the Algorithms A1 and A2 shall return optimum solutions with the same total weight. And the test results justify this conjecture. This value is used as the baseline, $W_{optimum}$ for the weight error rate calculation.

The general accuracy of the algorithms can be listed below from the best to the worst:

1. Algorithm A1 MWIS
2. Algorithm A2 AMISL (same as Algorithm MWIS)
3. Algorithm A5 MWIS_SubCS_GWMIN
4. Algorithm A8 MWIS_SubCS_GWMIN2
5. Algorithm A4 MWIS_CS_GWMIN
6. Algorithm A3 GWMIN
7. Algorithm A7 MWIS_CS_GWMIN2
8. Algorithm A6 GWMIN2

As listed above, merging the approximation algorithms with Algorithm A1 structure can improve the accuracy. And the test results justify the statement that applying the approximation algorithm on smaller subgraphs can achieve better accuracy, e.g., Algorithm A5 and A8 have better accuracy than the Algorithm A4 and A7, respectively.

# 8 Conclusions

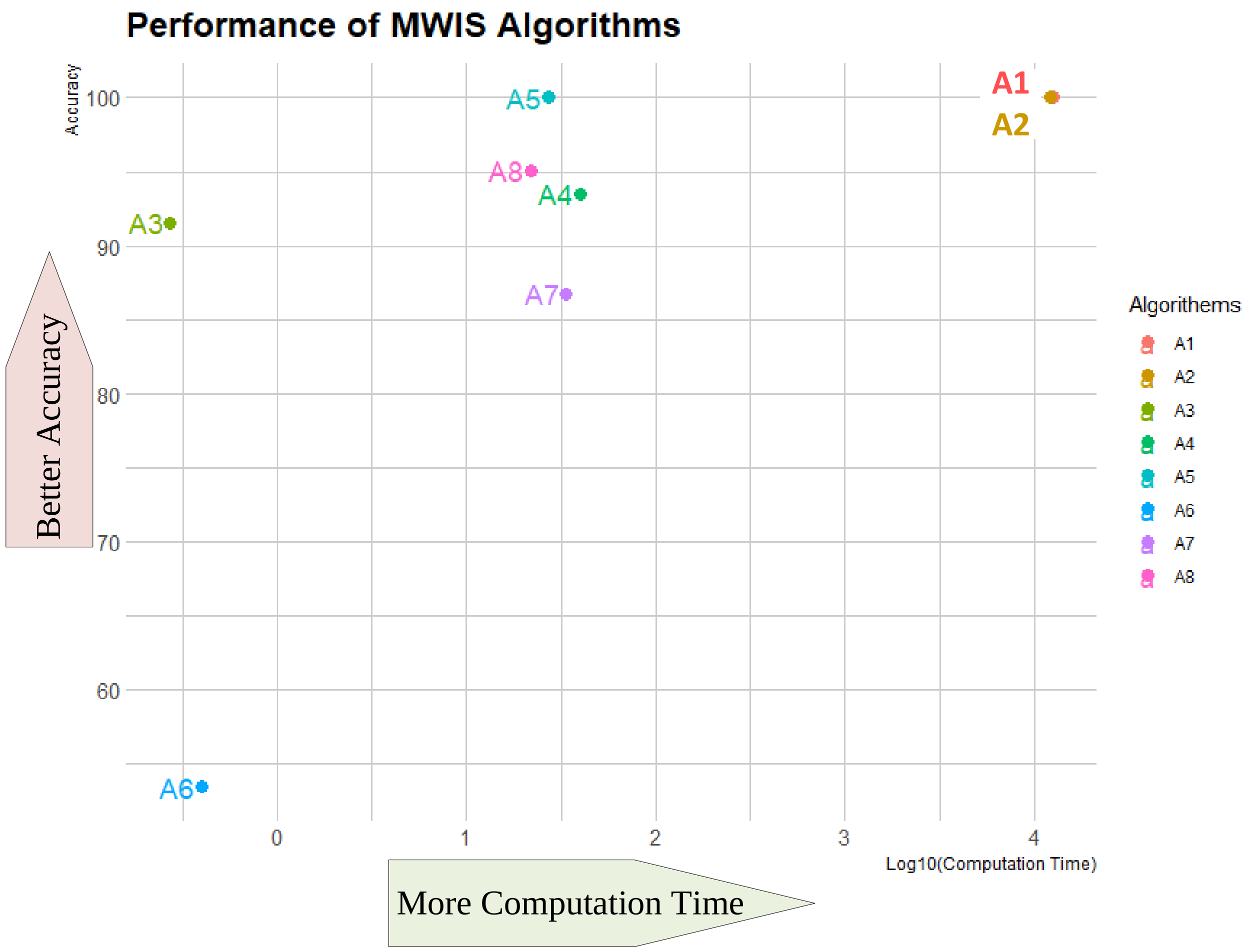

Figure 16. Performance of the MWIS Algorithms

This paper considers the MWIS problem on general graphs and develops the NHHA framework for solving the MWIS problem. In order to reduce the computational complexity, utility functions are developed or adopted; they are Algorithm 1: the basic cycles algorithm (Paton, 1969), Algorithm 2, the diameter algorithm (Takes & Kosters, 2011, 2013; Borassi et al., 2015), and Algorithm 3: the middle node algorithm. Using the NHHA framework, we establish that we always obtain feasible solutions to the MWIS problem. Two exact MWIS algorithms, Algorithm A1 MWIS and Algorithm A2 AMISL, are developed. For all the general graphs we have tested, solutions of the Algorithm A1 and A2 are always optimum.

Moreover, based on the NHHA framework, we develop four composed algorithms using the two merging methods with two approximation MWIS algorithms from literature, GWMIN and GWMIN2 (Sakai et al., 2003). Composing approximation algorithms with the NHHA framework is an effective way to improve the accuracy of approximation MWIS algorithms or the computational speed of Algorithm A1 MWIS. Composed algorithms, which are using the merging Method#2, are faster and better accuracy.

All eight algorithms for the MWIS problem, the exact MWIS algorithm, the AMISL algorithm, two approximation algorithms, and four composed algorithms, are applied and tested for solving the graph-based formulation of the resource-constrained PPS problem (Sun et al., preprint). The overall performance of the algorithms is illustrated in Figure 16. The general accuracy of the best five algorithms can be listed below from the best to the worst: Algorithm A1 MWIS; Algorithm A2 AMISL (same as Algorithm MWIS); Algorithm A5 MWIS_SubCS_GWMIN; Algorithm A8 MWIS_SubCS_GWMIN2; Algorithm A4 MWIS_CS_GWMIN. Note that all these algorithms considered satisfactory have the average error of less than 1% and the maximum error of less than 13% (The first four algorithms have the maximum error less than 9%) on all test instances.

In future work, it is of interest to investigate how to closely integrate the NHHA framework with application scenarios, in this case, the PPS problem. More specifically, utilizing both analytical and data-driven methods, how to utilize analytical and analytics methods for identifying features of the graph-based formulation of the PPS problem, how to fine-tune and closely-integrate the strategies of removed node selections, specialized approximation algorithms, and weight factor arrangements can efficiently improve MWIS algorithms with applications. Further on, we wish to explore the real-world applications with the NHHA framework and our graph-based problem formulation.

## Appendix I: An Example for Algorithm A1 on a Simple Graph

The exact MWIS algorithms described in Section 5 is complex. In Appendix I, we walk through Algorithm A1 in detail with a simple example in Figure A1. A simple weighted graph $G$ shown in Figure A1 is given, with the nodes, edges, and weights shown as Figure A1. Note that Algorithm A2 follows a similar process, but it is returning the AMIS at each step.

All the step indexes used below are from Algorithm A1.

In step (1.1), we need to perform step (1.1.1) to (1.1.5) to find and remove nodes and update the subgraphs dictionary (SD) accordingly:

$$SD\text{: \{the } removal\ node\text{: node sets of each } connected\ component\}$$

The SD is in the format that each node removed (removed node) is the key, and node sets of each connected component in the induced subgraphs are the value of the key. The node removal process iterates until the induced subgraphs satisfy the Theorem 1 conditions.

Perform step (1.1.1), to find the first removed node from the input graph; we need to find a cycle basis set of the input graph. Count the occurrence of each node in the cycle basis set; the first removed node is the node that has the most occurrences. Apply Algorithm 1, the cycle basis algorithm, to find a cycle basis set and count the number of cycles each node belongs to. The nodes and their counts are saved in a dictionary, “occurrence_dict”: {'1': 3, '0': 3', '3': 2, '4':

2, 2': 1, '5': 1, '6': 0, '7': 0, '8': 0, '9': 0, '10': 0, '11': 0}. The occurrence of node ‘1’ and node ‘0’ both are 3; we randomly pick node ‘1’ among them. Remove node ‘1’ and the adjunct edges, the induced subgraph is illustrated as Figure A2.

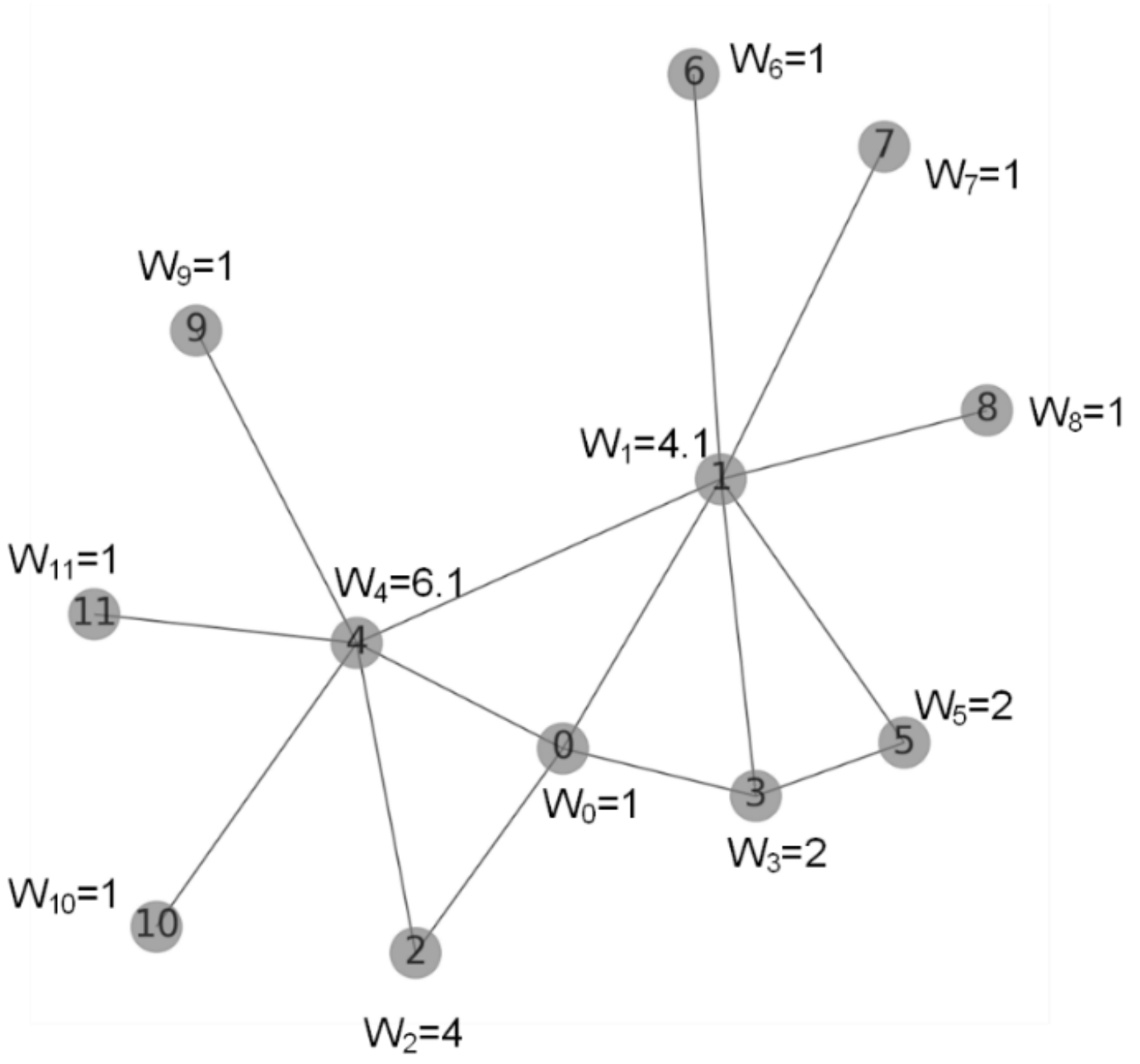

Figure A1. Simple graph for algorithm walk-through

Perform step (1.1.2), update SD with the key-value pair, SD: {'1': [{'6'}, {'7'}, {'8'}, {'4', '2', '10', '9', '11', '5', '0', '3'}]}. After removing the node ‘1’, the induced subgraph has four connected components, we use the node sets to denote these components, they are {'6'}, {'7'}, {'8'}, and {'4', '2', '10', '9', '11', '5', '0', '3'}.

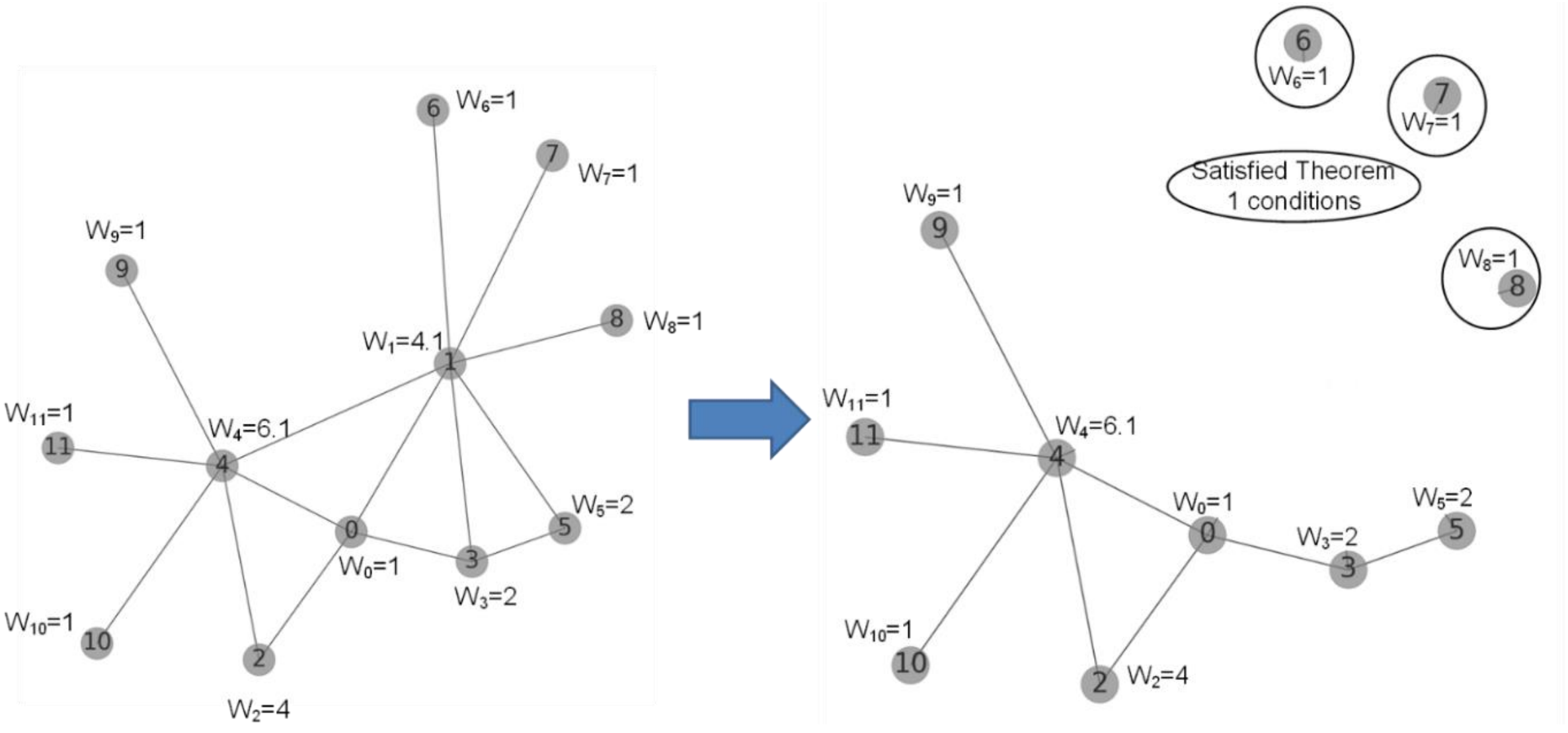

Figure A2a (left) & A2b (right). Remove Node ‘1’ from the Graph and the Induced Subgraph

Preform step (1.1.3), for each connected subgraph, exam whether they satisfy the Theorem 1 conditions. Among the four components, {'6'}, {'7'}, {'8'} satisfy the Theorem 1 conditions (in Figure A2b), but {'4', '2', '10', '9', '11', '5', '0', '3'} does not.

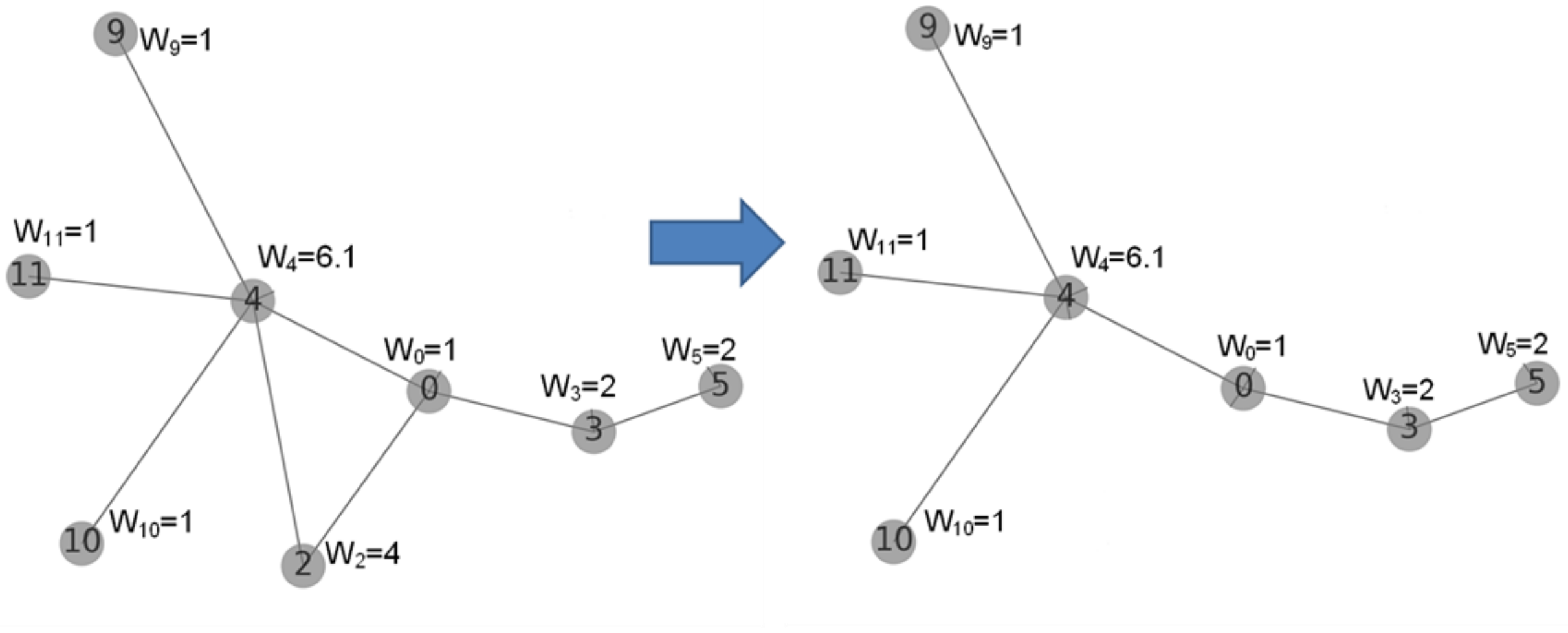

Figure A3a (left) & A3b (right). Remove Node ‘2’ and the Induced Subgraph

Preform step (1.1.4), the component subgraph, {'4', '2', '10', '9', '11', '5', '0', '3'}, does not satisfy the Theorem 1 conditions. Preform step (1.1.1), with the subgraph {'4', '2', '10', '9', '11', '5', '0', '3'}, apply Algorithm 1 to get current “occurrence_dict”: {'2': 1, '4': 1, '0': 1, '9': 0, '10': 0, '11': 0, '3': 0, '5': 0}. The occurrence of node ‘2’, node ‘4’ and node ‘0’ are 1, we randomly pick node ‘2’ among them. Remove node ‘2’ and the adjunct edges, the induced subgraph is illustrated as the Figure A3b.

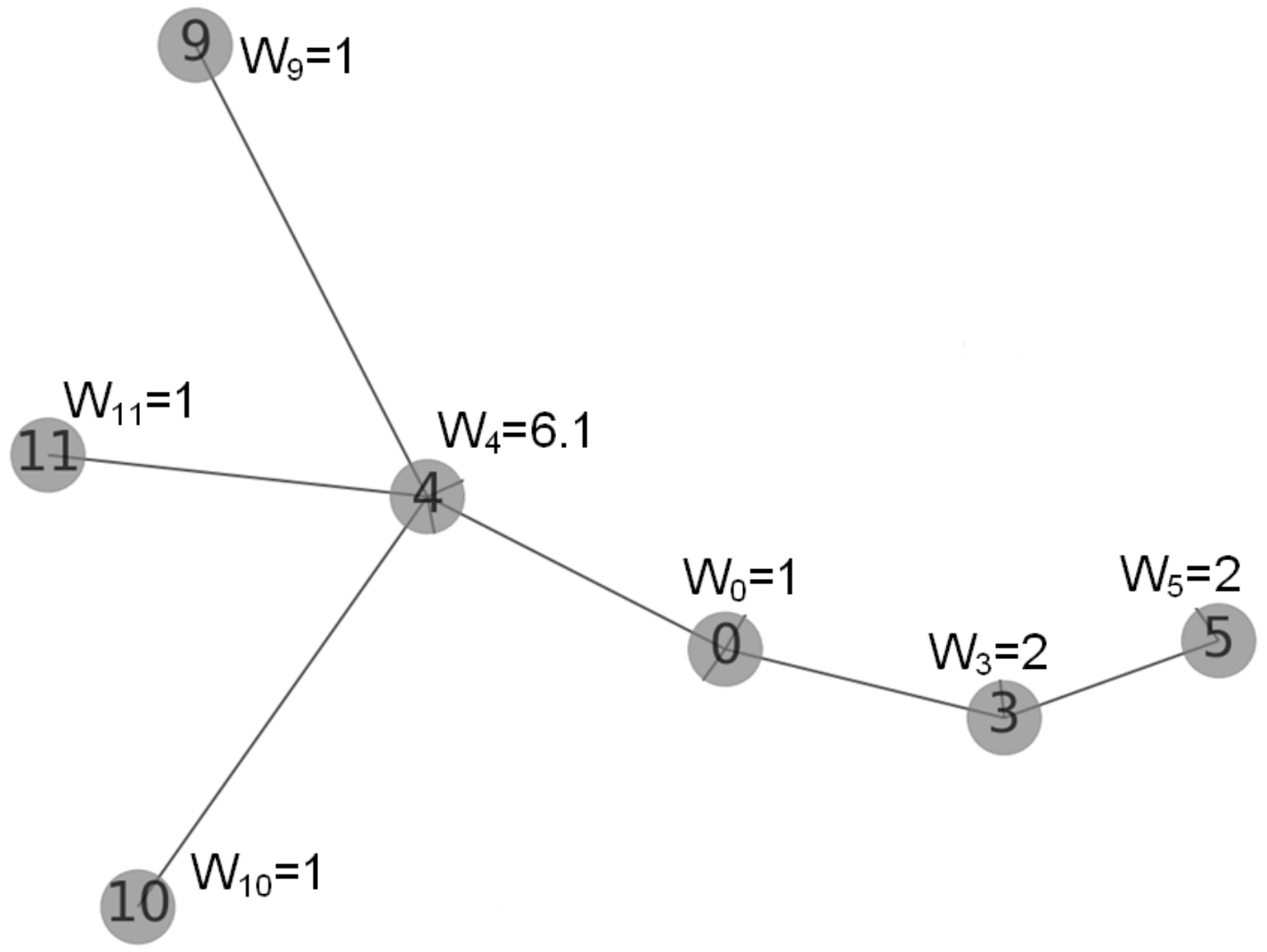

Figure A4. The induced subgraph after removing node ‘2’

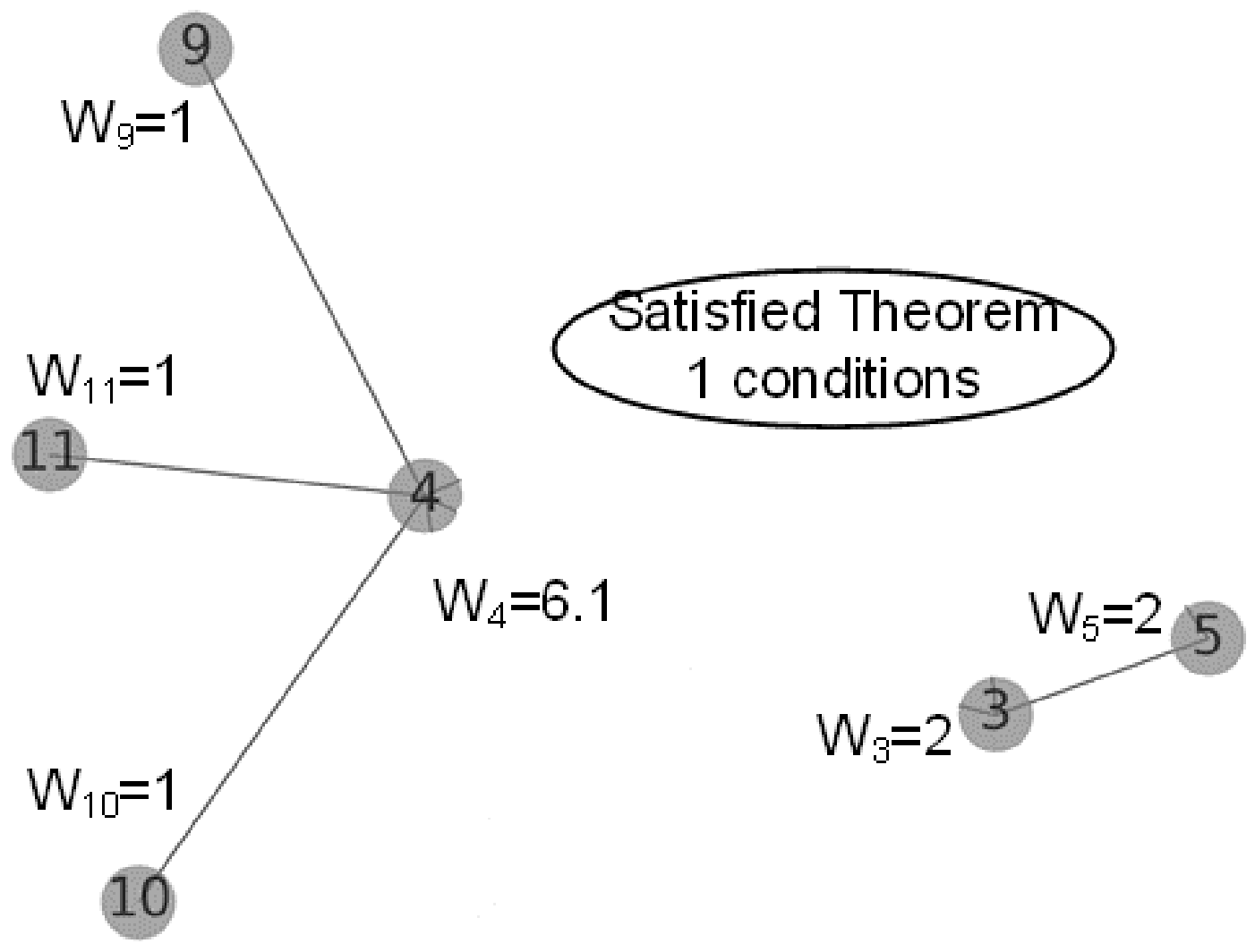

Figure A5. The induced subgraph after removing node ‘0’

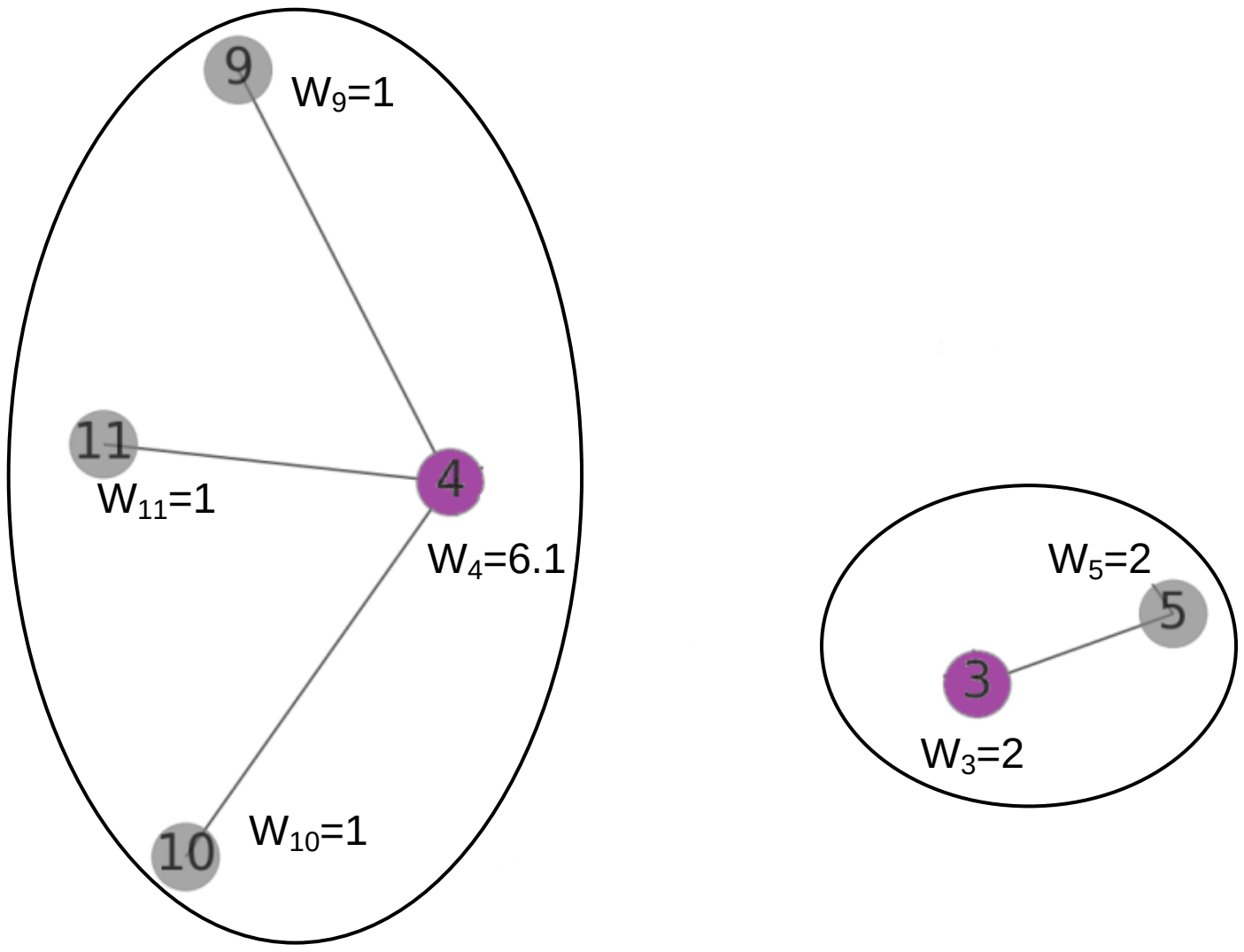

Figure A6. The Preliminary Set at the level node ‘0’

Preform step (1.1.2), update SD with the key-value pair, SD: {'1': [{'6'}, {'7'}, {'8'}, {'4', '2', '10', '9', '11', '5', '0', '3'}], '2': [{'4', '10', '9', '11', '5', '0', '3'}]}. Shown as the Figure 4, the induced subgraph {'4', '10', '9', '11', '5', '0', '3'} is connected.

Preform step (1.1.3), to exam the induced subgraph. The induced subgraph {'4', '10', '9', '11', '5', '0', '3'} as-in Figure A4 does not satisfy the Theorem 1 conditions.

Perform step (1.1.4), by applying Algorithm 1, there is no cycle left in the graph {'4', '10', '9', '11', '5', '0', '3'}. Then, apply Algorithm 2, the diameter algorithm, this tree structure has a $diameter = 4$, which does not satisfy the Theorem 1 conditions. Go to step (1.1.1), with the graph {'4', '10', '9', '11', '5', '0', '3'}, apply Algorithm 3, the middle node algorithm, to get the middle node ‘0’ of the tree. Remove node ‘0’ and the adjunct edges, the induced subgraph is illustrated as Figure A5.

Preform step (1.1.2), update SD with the key-value pair, SD: {'1': [{'6'}, {'7'}, {'8'}, {'9', '2', '5', '3', '4', '10', '11', '0'}], '2': [{'9', '5', '3', '4', '10', '11', '0'}], '0': [{'5', '3'}, {'10', '9', '11', '4'}]}. After removing the node ‘0’, the induced subgraph has two connected components, they are {'5', '3'}, and {'10', '9', '11', '4'}.

Perform step (1.1.3), to exam the induced subgraph. According to step (1.1.4), the two connected components in the induced subgraph both satisfy the Theorem 1 conditions shown in Figure A6.

Jump to step (1.1.5), when all subgraphs satisfy Theorem 1 conditions, return the latest SD: {'1': [{'6'}, {'7'}, {'8'}, {'9', '2', '5', '3', '4', '10', '11', '0'}], '2': [{'9', '5', '3', '4', '10', '11', '0'}], '0': [{'5', '3'}, {'10', '9', '11', '4'}]}.

All the procedures in step (1.1) for node removal are finished here.

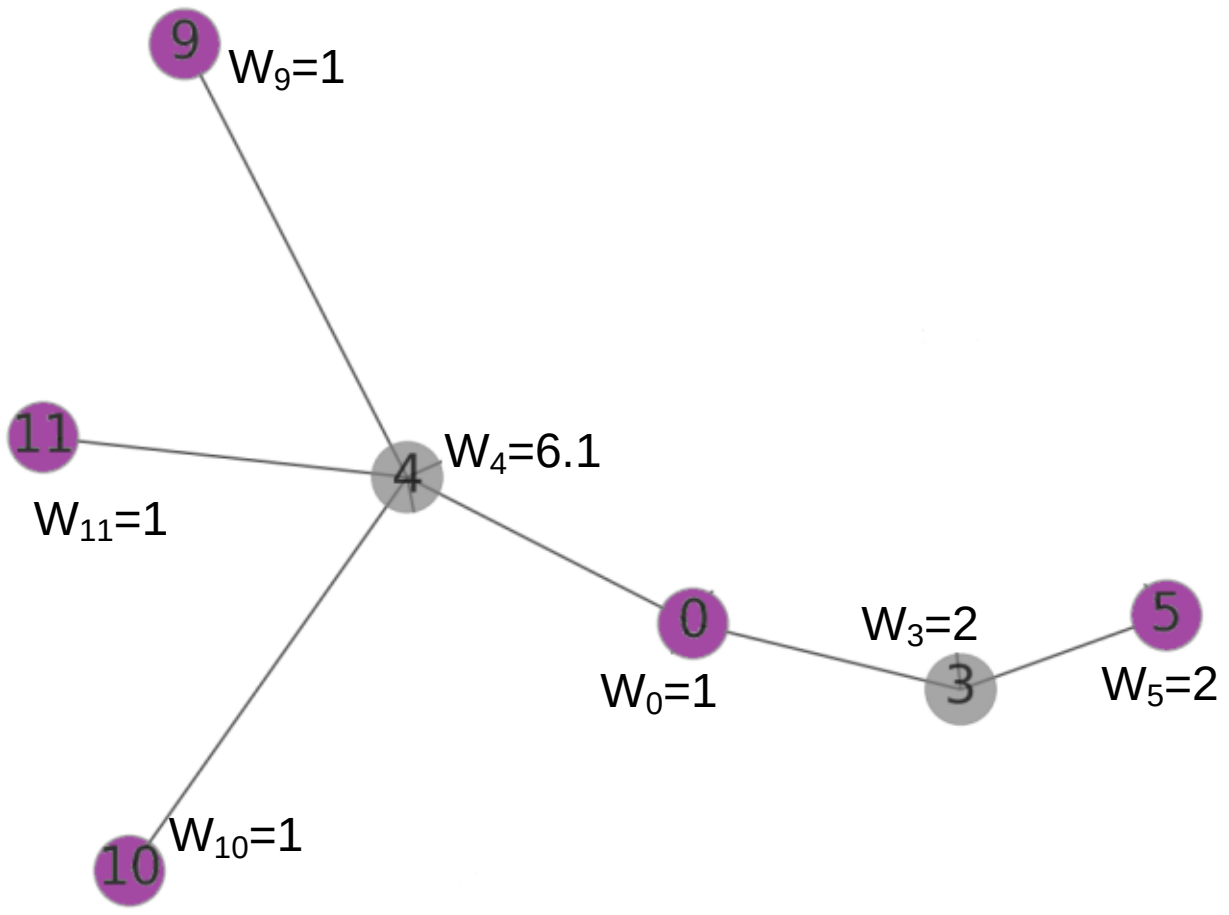

Figure A7. The Compare Set at the level node ‘0’

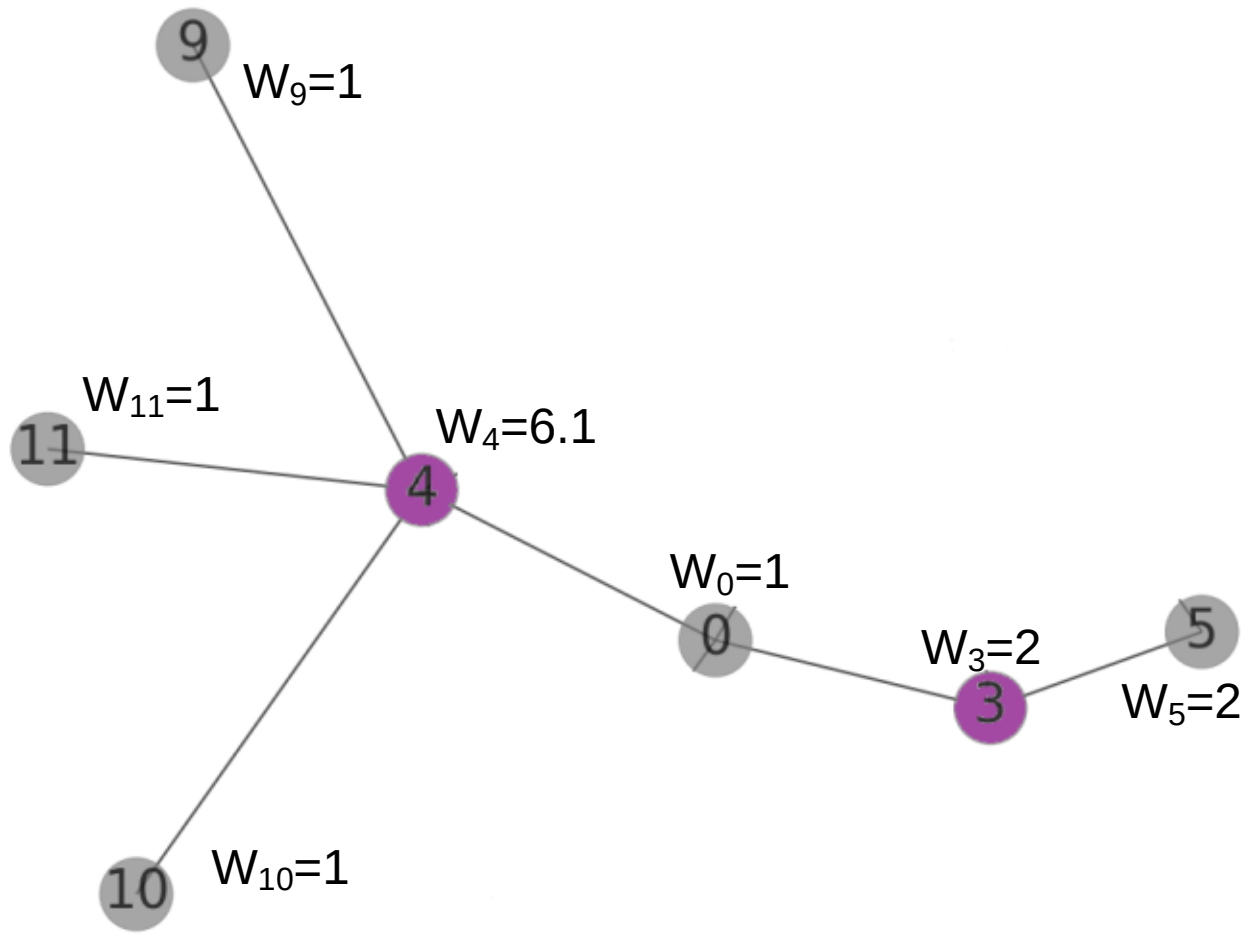

Figure A8. The MWIS at the level node ‘0’

Preform step (1.2), get the Preliminary Set from the induce subgraph according to the last key-value pair in SD. The last key-value pair in SD is: {'0': [{'5', '3'}, {'10', '9', '11', '4'}]}, indicating that at the level of node ‘0’, there are two connected components {'5', '3'} and {'10', '9', '11', '4'}. And the Theorem 1 conditions are satisfied. According to Theorem 1, we can find the Preliminary Set for the induced subgraph with nodes {'5', '3', '10', '9', '11', '4'}. This induced subgraph is called the Preliminary Set Subgraph (PSS) at level node ‘0’. The Preliminary Set at the level node ‘0’ is {4,3} with a weight total 8.1, shown as Figure A6.

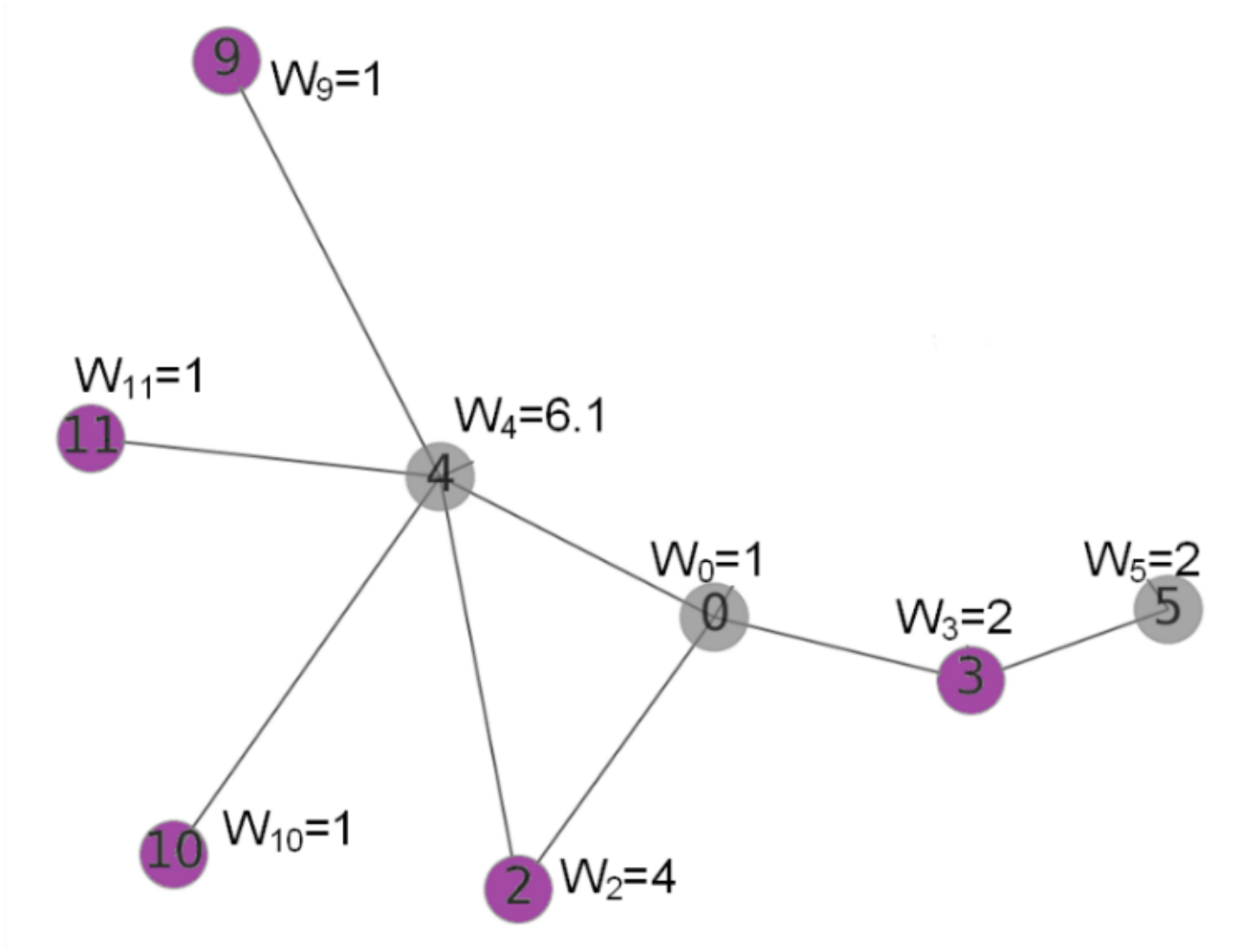

Figure A9. The Compare Set at the level node ‘2’

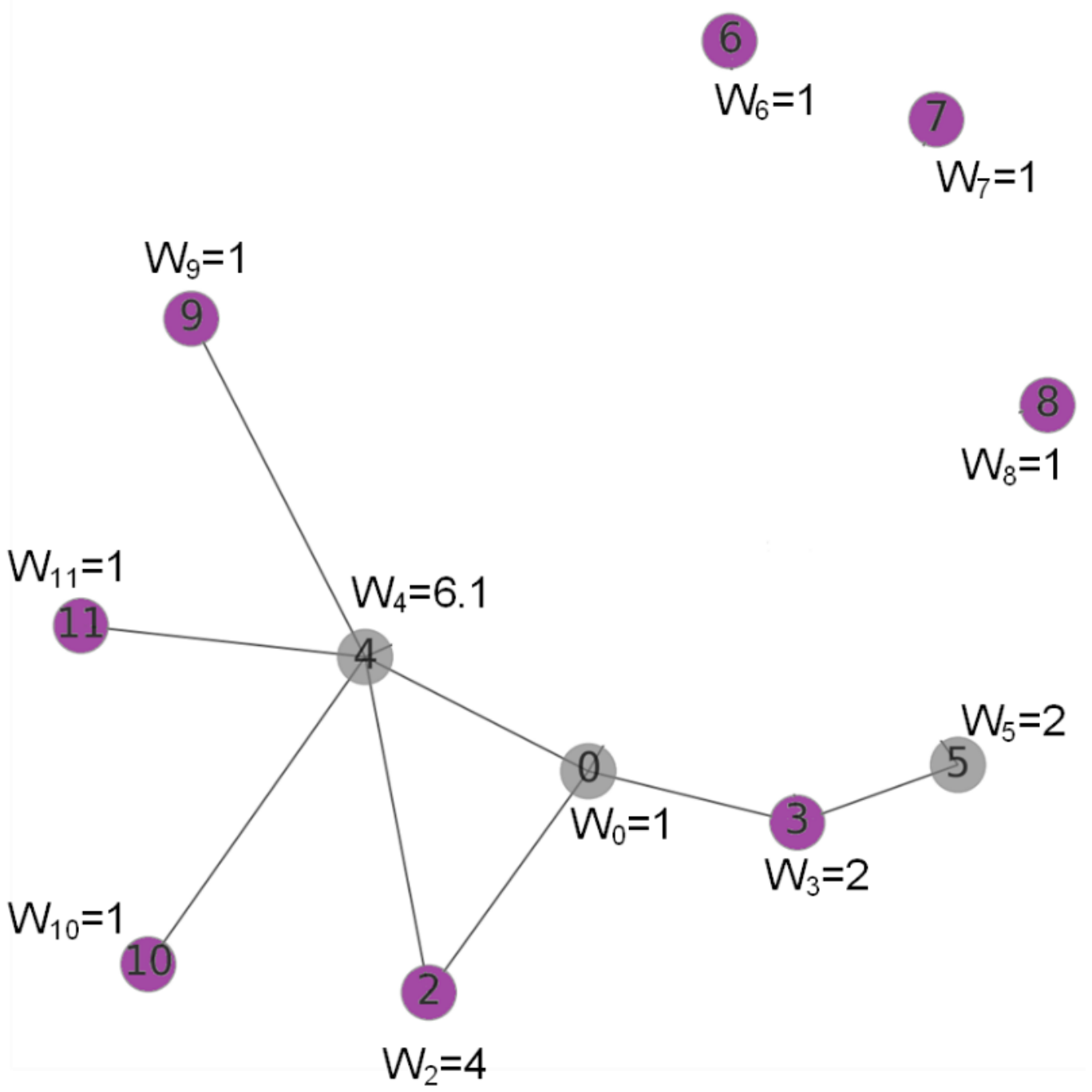

Figure A10. The Preliminary Set at the level node ‘1’

Perform step (1.3), the ‘last key’ is node ‘0’. Add node ‘0’ to the induced subgraph (Figure A6) of step (1.2), the induced graph rolls back to Figure A4 or Figure A7. Then, follow the adding node heuristics to find the Compare Set at level node ‘0’. Remove the neighbors of node ‘0’ and the adjacent edges of node ‘0’ from Figure A7, this induced subgraph with nodes {‘9’, ‘5’, ‘10’, ‘11’, ‘0’} is the Compare Set Subgraph (CSS) at the level removed node ‘0’. The Compare Set at the level node ‘0’ is {'5', '0', '11', '10', ‘9’} with a weight total 6, shown as Figure A7.

Perform step (1.4), according to Theorem 2, get the set with maximum weighted total among the two sets: the Preliminary Set (Figure A6) and the Compare Set (Figure A7) at the level node '0'. The MWIS of the induced subgraph in Figure A8 with nodes {'0', '5', '3', '10', '9', '11', '4'} is {'4', '3'}. We can say that at level node '0', the Preliminary Set is {'4', '3'} with a total weight of 8.1 as shown in Figure A8.

Preform step (1.5), update the SD as {'1': [{'6'}, {'7'}, {'8'}, {'9', '2', '5', '3', '4', '10', '11', '0'}], '2': [{'9', '5', '3', '4', '10', '11', '0'}]}. $SD \neq \emptyset$, go to step (1.2). The PSS at level node '2' is the induced subgraph with nodes {'9', '5', '3', '4', '10', '11', '0'}. The Preliminary Set at level node '2' is getting based on the previous step. The Preliminary Set at level node '2' is the MWIS of the induced subgraph with node {'4', '10', '9', '11', '5', '0', '3'}, which is {'4', '3'}, and the total weights is 8.1. Preform step (1.3), the last key-value pair is the level node '2'. Get the Compare Set at level node '2', follow the adding node heuristics. Then, the induced graph rolls back to Figure A3a. The CSS at level node '2' is the induced subgraph with nodes {'9', '2', '5', '3', '10', '11'}. The Compare Set at level node '2' is {'2', '3', '9', '10', '11'}. And the total weight is 9, shown as Figure A9.

Preform step (1.4), since the total weight of the Compare Set is greater than that of the Preliminary Set at level node '2', according to Theorem 2, the induced subgraph with nodes: {'4', '2', '10', '9', '11', '5', '0', '3'} at level node '2' has its MWIS as {'2', '3', '9', '10', '11'}, the total weight is 9.

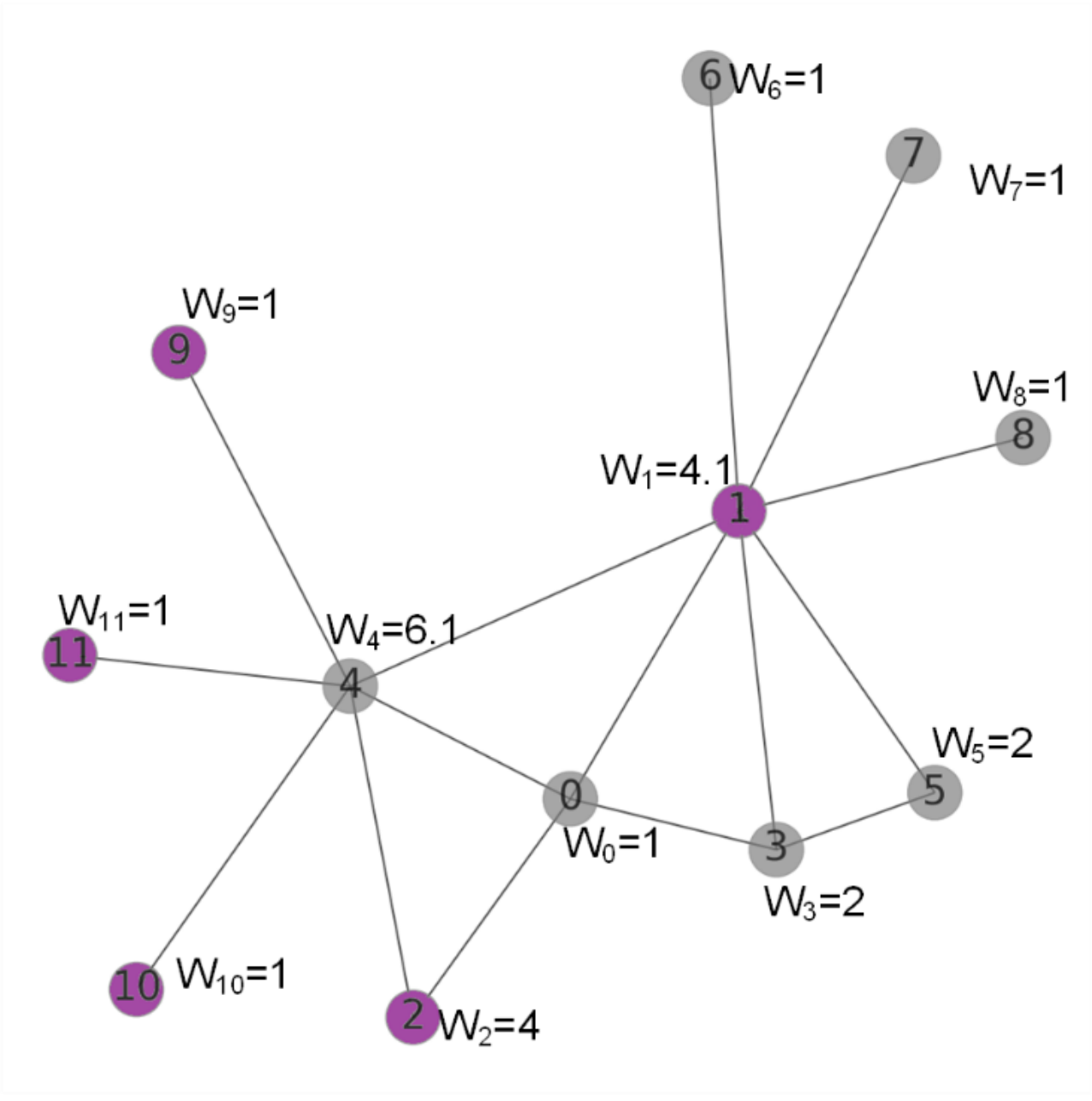

Figure A11. The Compare Set at the level node '1'

Perform step (1.5), update the SD as {'1': [{'6'}, {'7'}, {'8'}, {'9', '2', '5', '3', '4', '10', '11', '0'}]}. $SD \neq \emptyset$, go to step (1.2). The PSS at level node '1' is the induced subgraph with nodes {'9', '5', '3', '4', '10', '11', '0', '6', '7', '8'}. The Preliminary Set at level node '1' is based on the induce subgraph in the previous step. The induced subgraph at level node '1' has four components: {'6'}, {'7'}, {'8'}, and {'9', '2', '5', '3', '4', '10', '11', '0'}. For the connected components, the induced subgraph with nodes {'4', '2', '10', '9', '11', '5', '0', '3'}, has its MWIS as {'2', '3', '9', '10', '11'}, the total weight is 9, same as the MWIS as level node '2'. According to the Theorem 1 and Corollary 1, the Preliminary Set at level node '1' is the union of the MWIS of the four components with the node sets: {'6'}, {'7'}, {'8'}, and {'9', '2', '5', '3', '4', '10', '11', '0'}. The Preliminary Set at level node ‘1’ is $\{‘6’\} \cup \{‘7’\} \cup \{‘8’\} \cup \{'2','3','9','10','11'\}$, which has a total weight of 12, shown as Figure A10. Perform step (1.3), the last key-value pair is the level node ‘1’. Get the Compare Set at level node ‘1’ follow the adding node heuristics. Then, the induced graph rolls back to Figure A2a. The CSS at level node ‘1’ is the induced subgraph with nodes {'1', '9', '2', '10', '11'}. The Compare Set at level node ‘1’ is {'1', '2', '9', '10', '11’} with a total weight of 11.1, shown as Figure A11. Perform step (1.4), since the total weight of Preliminary Set is greater than that of Compare Set at level node ‘1’, according to Theorem 3-2, the induced subgraph with nodes: {'1', '6', '7', '8', '9', '2', '5', '3', '4', '10', '11', '0'} at level ‘1’ has its MWIS as {'6', '7', '8', '2', '3', '9', '10', '11’} the total weight is 12.

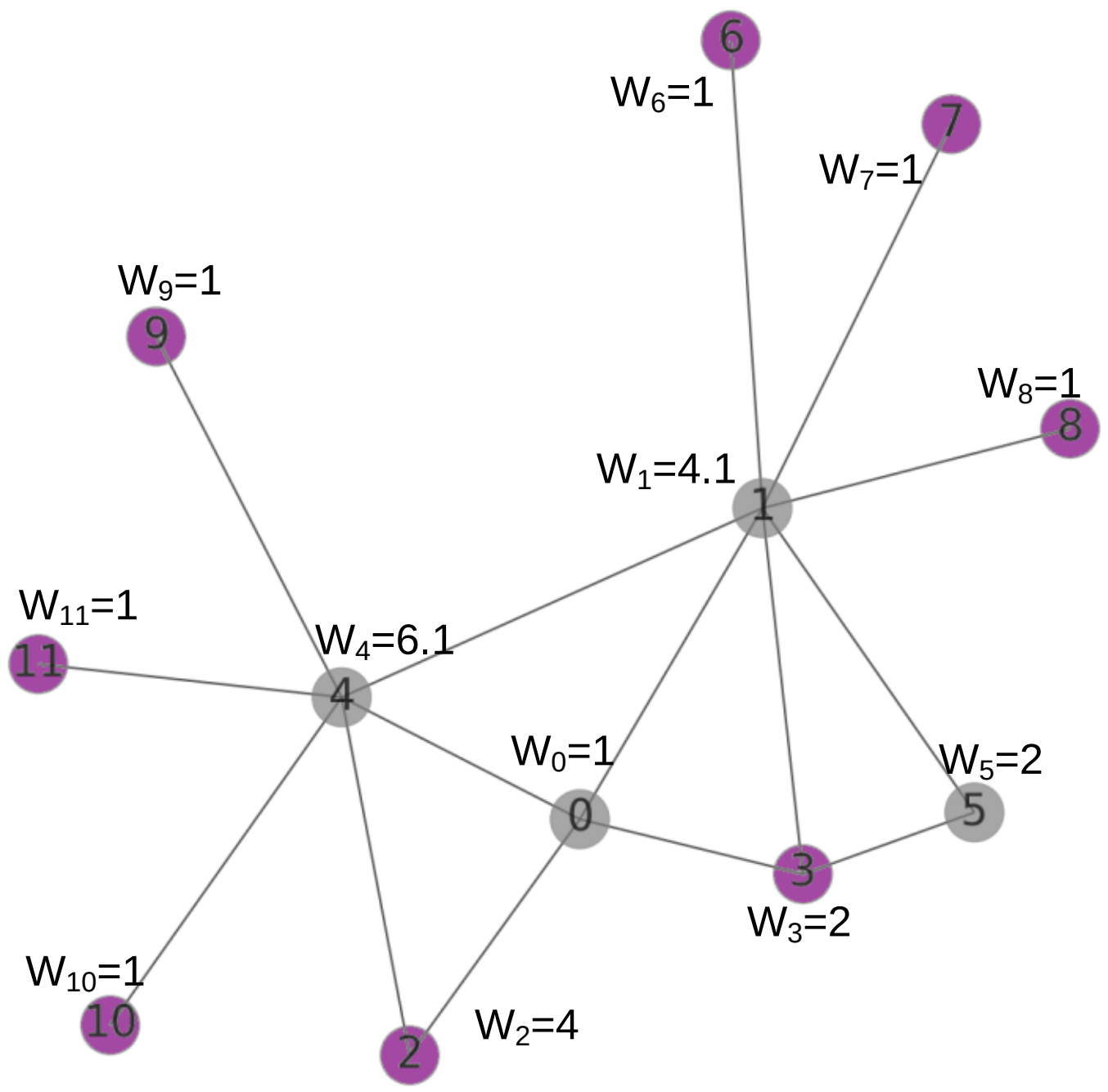

Figure A12. The MWIS of graph $G$

Perform step (1.5), update the SD, $SD = \emptyset$, return the MWIS of the original graph $G$. The MWIS is {'6', '7', '8', '2', '3', '9', '10', '11’}, the total weight is 12, shown as Figure A12.

# Appendix II: Test Details of MWIS Algorithms

| Test Info | | | | A1 | | A2 | | A3 | | A4 | | A5 | | A6 | | A7 | | A8 | |
|---|---|---|---|---|---|---|---|---|---|---|---|---|---|---|---|---|---|---|---|
| Test-ID | # of Edges | # of Nodes | Graph Density | Run Time | Weight Sum | Run Time | Weight Sum | Run Time | Weight Sum | Run Time | Weight Sum | Run Time | Weight Sum | Run Time | Weight Sum | Run Time | Weight Sum | Run Time | Weight Sum |
| 1 | 6 | 5 | 0.6 | 0.118567 | 0.00103 | 0.111136 | 0.00103 | 0.000253 | 0.00101 | 0.10485 | 0.00103 | 0.10639 | 0.00103 | 0.000239 | 0.00101 | 0.162926 | 0.00103 | 0.101249 | 0.00103 |
| 2 | 9 | 5 | 0.9 | 0.107675 | 0.25001 | 0.087636 | 0.25001 | 0.000233 | 0.25 | 0.103948 | 0.25001 | 0.10591 | 0.25001 | 0.000207 | 0.25 | 0.106247 | 0.25001 | 0.106666 | 0.25001 |
| 3 | 29 | 12 | 0.439393939 | 0.393995 | 4.666676667 | 0.273826 | 4.666676667 | 0.000623 | 4.666676667 | 0.287531 | 4.666676667 | 0.293275 | 4.666676667 | 0.000625 | 4.583333333 | 0.253069 | 4.666676667 | 0.253618 | 4.666676667 |
| 4 | 29 | 14 | 0.318681319 | 0.485537 | 7.75001 | 0.355922 | 7.75001 | 0.001504 | 7.75001 | 0.292077 | 7.75001 | 0.382857 | 7.75001 | 0.000786 | 7.583333333 | 0.306305 | 7.75001 | 0.290889 | 7.75001 |
| 5 | 41 | 17 | 0.301470588 | 1.748986 | 3.00727 | 1.641865 | 3.00727 | 0.001492 | 3.00727 | 0.713053 | 3.00727 | 0.688541 | 3.00727 | 0.001111 | 2.090593333 | 0.345433 | 2.75729 | 0.397687 | 2.75729 |
| 6 | 44 | 17 | 0.323529412 | 0.729959 | 17.04168667 | 0.701199 | 17.04168667 | 0.002286 | 17.04168667 | 0.600235 | 17.04168667 | 0.344752 | 17.04168667 | 0.001547 | 16.45834333 | 0.469231 | 17.04168667 | 0.321118 | 17.04168667 |
| 7 | 69 | 17 | 0.507352941 | 0.561269 | 4.25001 | 0.368487 | 4.25001 | 0.001185 | 4.25 | 0.403378 | 4.25001 | 0.427379 | 4.25001 | 0.00099 | 4.25 | 0.405091 | 4.25001 | 0.409639 | 4.25001 |
| 8 | 52 | 18 | 0.339869281 | 1.448404 | 1.340603333 | 0.94847 | 1.340603333 | 0.002395 | 1.340603333 | 0.811353 | 1.340603333 | 0.627162 | 1.340603333 | 0.001794 | 0.590593333 | 0.571024 | 1.340603333 | 0.569017 | 1.340603333 |
| 9 | 6 | 19 | 0.035087719 | 0.556402 | 89.50005 | 0.560922 | 89.50005 | 0.002827 | 89.50005 | 0.307328 | 89.50005 | 0.310983 | 89.50005 | 0.001498 | 87.25004 | 0.424144 | 89.50005 | 0.324134 | 89.50005 |
| 10 | 0 | 20 | 0 | 0.001093 | 1639.152818 | 0.00279 | 1639.152818 | 0.002863 | 1639.152818 | 0.001087 | 1639.152818 | 0.001054 | 1639.152818 | 0.002053 | 1639.152818 | 0.001059 | 1639.152818 | 0.001054 | 1639.152818 |
| 11 | 8 | 25 | 0.026666667 | 2.112017 | 164.5556056 | 1.167152 | 164.5556056 | 0.005565 | 164.5556056 | 0.510531 | 164.5556056 | 0.503593 | 164.5556056 | 0.00294 | 164.5556056 | 0.723653 | 164.5556056 | 0.741895 | 164.5556056 |
| 12 | 50 | 26 | 0.153846154 | 1.581559 | 19.07118667 | 1.164716 | 19.07118667 | 0.006769 | 19.07118667 | 0.75344 | 19.07118667 | 0.716165 | 19.07118667 | 0.001923 | 17.73784333 | 0.756383 | 19.07118667 | 0.719825 | 19.07118667 |
| 13 | 101 | 26 | 0.310769231 | 5.463803 | 10.51577 | 4.053637 | 10.51577 | 0.004824 | 10.51577 | 0.58007 | 10.51577 | 1.428015 | 10.51577 | 0.002125 | 9.01578 | 0.642795 | 10.51577 | 0.641116 | 10.51577 |
| 14 | 115 | 27 | 0.327635328 | 0.401543 | 839.5416767 | 0.242558 | 839.5416767 | 0.008301 | 839.5416767 | 0.271618 | 839.5416767 | 0.194913 | 839.5416767 | 0.00288 | 839.5416767 | 0.26215 | 839.5416767 | 0.192846 | 839.5416767 |
| 15 | 147 | 28 | 0.388888889 | 2.431524 | 40.02782778 | 2.351423 | 40.02782778 | 0.007439 | 40.02782778 | 0.605325 | 40.02782778 | 0.507716 | 40.02782778 | 0.002718 | 39.11115111 | 0.790927 | 39.11115111 | 0.524695 | 39.11115111 |
| 16 | 155 | 30 | 0.356321839 | 2.160428 | 14.77777 | 1.611405 | 14.77777 | 0.004921 | 14.77777 | 0.935768 | 14.77777 | 0.821684 | 14.77777 | 0.003878 | 13.52776 | 0.975734 | 14.77777 | 0.857362 | 14.77777 |
| 17 | 135 | 31 | 0.290322581 | 1.605572 | 31.37338333 | 1.254355 | 31.37338333 | 0.008557 | 31.37338333 | 1.059978 | 31.37338333 | 0.697107 | 31.37338333 | 0.003946 | 29.54004 | 0.804118 | 31.37338333 | 0.704573 | 31.37338333 |
| 18 | 198 | 31 | 0.425806452 | 2.601461 | 549.3889089 | 2.101934 | 549.3889089 | 0.010304 | 547.8055656 | 0.723661 | 549.3889089 | 0.678604 | 549.3889089 | 0.002738 | 310.9444444 | 0.693808 | 549.3889089 | 0.777898 | 549.3889089 |
| 19 | 325 | 34 | 0.579322638 | 6.154325 | 43.13892889 | 4.924645 | 43.13892889 | 0.007267 | 43.13892889 | 1.771377 | 43.13892889 | 0.90644 | 43.13892889 | 0.003798 | 12.02777778 | 1.087117 | 41.72225222 | 0.761185 | 41.72225222 |
| 20 | 216 | 36 | 0.342857143 | 2.129672 | 13.87584333 | 1.658275 | 13.87584333 | 0.010086 | 13.87584333 | 0.818565 | 13.87584333 | 0.419063 | 13.87584333 | 0.003445 | 13.87584333 | 0.49319 | 13.87584333 | 0.367221 | 13.87584333 |
| 21 | 264 | 39 | 0.356275304 | 10.61476 | 88.92681778 | 9.012758 | 88.92681778 | 0.014521 | 88.92681778 | 3.646862 | 88.92681778 | 0.890331 | 88.92681778 | 0.004023 | 54.19858556 | 1.00701 | 88.92681778 | 0.841288 | 88.92681778 |
| 22 | 313 | 39 | 0.422402159 | 3.862964 | 39.62565111 | 3.050283 | 39.62565111 | 0.015739 | 39.13094333 | 1.032787 | 39.13094333 | 0.960322 | 39.62565111 | 0.006402 | 36.81481778 | 1.01599 | 39.13094333 | 0.954846 | 39.62565111 |
| 23 | 336 | 43 | 0.372093023 | 3.418986 | 155.50004 | 2.797809 | 155.50004 | 0.022145 | 155.50004 | 1.047748 | 155.50004 | 0.40617 | 155.50004 | 0.008996 | 105.5833633 | 0.457707 | 155.50004 | 0.356102 | 155.50004 |
| 24 | 388 | 45 | 0.391919192 | 2.597778 | 1309.291677 | 2.368349 | 1309.291677 | 0.019715 | 1309.291677 | 1.403225 | 1309.291677 | 0.517649 | 1309.291677 | 0.007383 | 1309.291677 | 0.765621 | 1309.291677 | 0.507685 | 1309.291677 |
| 25 | 425 | 46 | 0.410628019 | 8.003977 | 70.91673667 | 7.308244 | 70.91673667 | 0.02527 | 70.91673667 | 1.563586 | 70.91673667 | 0.675255 | 70.91673667 | 0.007505 | 55.66670667 | 0.702882 | 70.91673667 | 0.602179 | 70.91673667 |
| 26 | 583 | 53 | 0.423076923 | 18.92629 | 76.21324667 | 15.78386 | 76.21324667 | 0.033673 | 76.21324667 | 4.760113 | 76.21324667 | 1.667401 | 76.21324667 | 0.028344 | 73.71323667 | 2.824472 | 73.71323667 | 1.518027 | 73.71323667 |
| 27 | 465 | 54 | 0.324947589 | 9.57828 | 50.72217667 | 8.292436 | 50.72217667 | 0.015988 | 44.54351 | 1.105264 | 45.38884333 | 0.605086 | 50.72217667 | 0.018954 | 45.38884333 | 0.955258 | 45.38884333 | 0.50795 | 50.72217667 |
| 28 | 707 | 56 | 0.459090909 | 15.62693 | 836.4167067 | 12.78697 | 836.4167067 | 0.034683 | 836.4167067 | 2.639332 | 836.4167067 | 1.155839 | 836.4167067 | 0.012823 | 453.6388989 | 1.231597 | 836.4167067 | 1.106137 | 836.4167067 |
| 29 | 726 | 59 | 0.424313267 | 12.17281 | 146.37504 | 10.34992 | 146.37504 | 0.036667 | 124.0417067 | 1.574886 | 146.37504 | 1.028119 | 146.37504 | 0.014819 | 95.95840333 | 1.982005 | 124.0417067 | 0.982974 | 146.37504 |
| 30 | 726 | 59 | 0.424313267 | 11.86624 | 1510.37504 | 10.29127 | 1510.37504 | 0.033442 | 1457.041707 | 1.685658 | 1510.37504 | 1.029224 | 1510.37504 | 0.014331 | 1227.458403 | 1.065756 | 1457.041707 | 0.99773 | 1510.37504 |
| 31 | 723 | 61 | 0.395081967 | 25.16087 | 126.5310522 | 22.26743 | 126.5310522 | 0.085729 | 126.5310522 | 6.561642 | 126.5310522 | 2.307274 | 126.5310522 | 0.041315 | 123.3643756 | 0.176591 | 123.3643756 | 1.650675 | 123.3643756 |
| 32 | 840 | 63 | 0.430107527 | 45.89679 | 86.97227222 | 42.46756 | 86.97227222 | 0.051448 | 86.97227222 | 5.779079 | 86.97227222 | 2.353676 | 86.97227222 | 0.017616 | 63.97223222 | 3.711723 | 85.05559556 | 2.361078 | 85.05559556 |
| 33 | 980 | 67 | 0.443238354 | 16.00381 | 148.5972622 | 14.50707 | 148.5972622 | 0.054451 | 143.2639289 | 3.510512 | 148.5972622 | 1.769027 | 148.5972622 | 0.022087 | 98.51395889 | 1.745536 | 124.4861511 | 1.450741 | 148.5972622 |
| 34 | 1238 | 79 | 0.401817592 | 108.7613 | 146.75005 | 93.71785 | 146.75005 | 0.074164 | 146.75005 | 4.028621 | 146.75005 | 2.699751 | 146.75005 | 0.030016 | 38.77783778 | 2.930075 | 145.75005 | 2.488386 | 145.75005 |
| 35 | 1387 | 89 | 0.354187947 | 106.4795 | 237.4167367 | 94.94339 | 237.4167367 | 0.107995 | 237.4167367 | 4.876775 | 237.4167367 | 2.619159 | 237.4167367 | 0.050076 | 216.25009 | 2.955594 | 216.25009 | 2.067732 | 237.4167367 |
| 36 | 1387 | 89 | 0.354187947 | 104.5018 | 156.7474867 | 94.51976 | 156.7474867 | 0.094051 | 156.7474867 | 5.352133 | 156.7474867 | 2.602482 | 156.7474867 | 0.042201 | 91.73248667 | 2.735964 | 124.7384867 | 2.070066 | 156.7474867 |
| 37 | 1387 | 89 | 0.354187947 | 104.5936 | 2579.166737 | 95.38101 | 2579.166737 | 0.088363 | 2579.166737 | 4.599551 | 2579.166737 | 2.542564 | 2579.166737 | 0.041883 | 2087.916677 | 2.835716 | 2277.166737 | 2.062046 | 2579.166737 |
| 38 | 2184 | 110 | 0.364303586 | 460.6811 | 259.8334333 | 457.9928 | 259.8334333 | 0.146865 | 227.5834933 | 10.13942 | 227.5834933 | 5.899711 | 259.8334333 | 0.097114 | 241.3334933 | 7.294095 | 241.3334933 | 4.462351 | 254.0834333 |
| 39 | 2184 | 110 | 0.364303586 | 487.6465 | 2670.083433 | 484.6684 | 2670.083433 | 0.157923 | 2490.583493 | 9.573596 | 2658.916767 | 5.832959 | 2670.083433 | 0.081097 | 1988.083373 | 5.909477 | 2113.333483 | 4.697975 | 2431.916817 |
| 40 | 3108 | 126 | 0.394666667 | 978.4428 | 262.2778778 | 924.2892 | 262.2778778 | 0.22231 | 226.7223322 | 13.62323 | 262.2778778 | 9.455681 | 262.2778778 | 0.168643 | 201.6946044 | 11.76862 | 244.3334933 | 7.541868 | 262.2778778 |
| 41 | 3108 | 126 | 0.394666667 | 967.6664 | 2672.527878 | 927.3733 | 2672.527878 | 0.235084 | 2672.527878 | 17.15051 | 2672.527878 | 9.302291 | 2672.527878 | 0.176047 | 1885.527818 | 10.97006 | 2498.583493 | 7.475677 | 2672.527878 |
| 42 | 4718 | 161 | 0.366304348 | 13046.56 | 175.4459711 | 12705.98 | 175.4459711 | 0.274509 | 175.4459711 | 36.28218 | 175.4459711 | 28.67411 | 175.4459711 | 0.401469 | 90.89372333 | 34.98096 | 142.4607367 | 22.34016 | 170.5250444 |
| 43 | 4718 | 161 | 0.366304348 | 12041.21 | 260.1112211 | 11761.95 | 260.1112211 | 0.272659 | 260.1112211 | 43.85244 | 260.1112211 | 25.79004 | 260.1112211 | 0.395887 | 161.8890189 | 32.65466 | 202.8612811 | 21.44091 | 250.9445444 |